\documentclass[a4paper,11pt]{article}

\usepackage[utf8x]{inputenc}
\usepackage{color}

\newcommand{\Dx}{\boldsymbol{\nabla}}
\newcommand{\vel}{\boldsymbol{U}}
\newcommand{\grav}{\boldsymbol{g}}
\newcommand{\ez}{\boldsymbol{\hat{e}}_z}

\newcommand{\Ttrip}{T_{\mathrm{trip}}}
\newcommand{\Eov}{E_{0v}}
\newcommand{\pvstar}{p_v^*}
\newcommand{\ptrip}{p_{\mathrm{trip}}}
\newcommand{\cva}{c_{va}}
\newcommand{\cvv}{c_{vv}}
\newcommand{\cvl}{c_{vl}}
\newcommand{\cvm}{c_{vm}}
\newcommand{\cpa}{c_{pa}}
\newcommand{\cpv}{c_{pv}}

\newcommand{\cpm}{c_{pm}}
\newcommand{\qvstar}{q_v^*}
\newcommand{\ehat}{\widehat{e}}

\newcommand{\aav}{\alpha_v}
\newcommand{\bbv}{\beta_v}

\newcommand{\sfrac}[2]{\mathchoice
  {\kern0em\raise.5ex\hbox{\the\scriptfont0 #1}\kern-.15em/
   \kern-.15em\lower.25ex\hbox{\the\scriptfont0 #2}}
  {\kern0em\raise.5ex\hbox{\the\scriptfont0 #1}\kern-.15em/
   \kern-.15em\lower.25ex\hbox{\the\scriptfont0 #2}}
  {\kern0em\raise.5ex\hbox{\the\scriptscriptfont0 #1}\kern-.2em/
   \kern-.15em\lower.25ex\hbox{\the\scriptscriptfont0 #2}}
  {#1\!/#2}}
\newcommand{\hhat}{\widehat{h}}
\newcommand{\Hhat}{\mathcal{H}}
\newcommand{\x}{\boldsymbol{x}}
\newcommand{\po}{p_0}
\newcommand{\talpha}{\widetilde{\alpha}}
\newcommand{\tS}{\widetilde{S}}
\newcommand{\tgamma}{\widetilde{\gamma}_m}
\newcommand{\tsigma}{\widetilde{\sigma}}
\newcommand{\Ag}{\varPhi}
\newcommand{\gammabar}{\overline{\Gamma}_1}
\newcommand{\cdotb}{\boldsymbol{\cdot}}

\newcommand{\rhozero}{\rho_0}

\newcommand{\Aev}{{\mathcal{A}}_{e}}
\newcommand{\Bev}{{\mathcal{B}}_{e}}
\newcommand{\Cev}{{\mathcal{C}}_{e}}

\newcommand{\gammambar}{\overline{\gamma}_m}
\newcommand{\tgammambar}{\overline{\widetilde{\gamma}}_m}

\def\ds{\displaystyle}

\usepackage{graphicx}
\usepackage{amssymb}
\usepackage{amsmath}
\usepackage{enumerate}

\title{A Low Mach Number Model for Moist Atmospheric Flows}
\author{\textsc{Max~Duarte}\thanks{\textit{Corresponding author address:}
Max Duarte, Center for Computational Sciences and Engineering,
Lawrence Berkeley National Laboratory,
MS 50A-1148, 1 Cyclotron Rd. 
Berkeley, CA 94720
\newline{E-mail: MDGonzalez@lbl.gov}}
\textsc{, Ann~S.~Almgren, and John~B.~Bell}  \\
\textit{\footnotesize{CCSE, Lawrence Berkeley National Laboratory, Berkeley, California}}
\and
}
\begin{document}

\maketitle

\begin{abstract}
We introduce a low Mach number model for moist atmospheric
flows that accurately incorporates reversible moist processes 
in flows whose features of interest occur on advective rather 
than acoustic time scales.  
Total water is used as a prognostic variable, so that water vapor and 
liquid water are diagnostically recovered as needed
from an exact Clausius--Clapeyron formula for moist thermodynamics.
Low Mach number models can be computationally more efficient than 
a fully compressible model, but the low Mach number formulation
introduces additional mathematical and computational complexity 
because of the divergence constraint imposed on the velocity field.
Here, latent heat release is accounted for in the source term of
the constraint by estimating the rate of phase change based on
the time variation of saturated water vapor subject to 
the thermodynamic equilibrium constraint.
We numerically assess the validity of the low Mach number approximation 
for moist atmospheric flows by contrasting the low Mach number
solution to reference solutions computed with a fully compressible formulation
for a variety of test problems.
\end{abstract}

%%%%%%%%%%%%%%%%%%%%%%%%%%%%%%%%%%%%%%%%%%%%%%%%%%%%%%%%%%%%%%%%%%%%%

\section{Introduction}
Small--scale atmospheric phenomena are typically characterized by relatively 
slow dynamics, that is, low Mach number flows for which the fast acoustic modes
are physically irrelevant.  Thus, numerical modeling of these flows
does not typically require explicitly resolving fast--propagating sound waves.
Two different approaches have been widely used to remove the time step
constraint that would result from resolving fast acoustic modes.
The first and more common approach solves the fully compressible
equations of motion 
but limits the impact of acoustic modes, for instance, by advancing the acoustic 
signal in time with an implicit time discretization or with multiple smaller time steps,
as originally considered for cloud models in \cite{Tapp1976}
and \cite{Klemp1978}.  A second alternative consists 
of analytically filtering acoustic modes from the original compressible 
equations, thus deriving a new set of equations, often called 
sound--proof equations.  Within this category are
anelastic \cite{Bat53,OguPhi62,DutFic69,Gou69} 
and pseudo--incompressible \cite{durran:1989} models.

Several anelastic formulations (see, e.g., \cite{Clark1979,LipHem82,Grabowski2002}),
and recently a pseudo--incompressible formalism 
\cite{ONeill2013}, have been developed for moist flows.
In this paper we derive a low Mach number model for moist
atmospheric flows 
with a general equation of state
using the low Mach number formalism in \cite{ABNZ:III} 
as a starting point.   
Here we use the term ``low Mach number model'' to refer
to a model in which the equations are valid approximations to the
fully compressible equations when the Mach number is small.   
In atmospheric modeling, the equations which follow from
assuming the Mach number is small, thus the variation of the
pressure from the background pressure is small,
are also referred to as pseudo--incompressible equations
following \cite{durran:1989}.
The anelastic equations require additional assumptions on the
smallness of density and temperature variations to be valid. 
(For a complete discussion on sound-proof equations
for atmospheric flows,
and their differences, see,
for instance, \cite{Klein2009} and references therein.)
We note that the low Mach number equations presented here do
not guarantee that a flow that initially satisfies the low Mach
number assumption will continue to satisfy it for all time.
The buoyancy forcing from a large density perturbation in a 
domain with large vertical extent could accelerate 
the flow into a regime where the Mach number is no longer small.  
Once the flow reached this regime, the low Mach number equations 
would no longer be valid approximations to the fully compressible 
equations.  However, until that point, the equations are valid.

Analogous to 
the moist pseudo--incompressible model of
\cite{ONeill2013} we consider only reversible
processes given by water phase changes, using here an exact
Clausius--Clapeyron formula for moist thermodynamics.
In contrast to \cite{ONeill2013}, however, 
we derive the equations of motion in terms of conserved variables
(like \cite{Ooyama1990} in a compressible framework),
that is, using appropriate invariant variables such that terms 
resulting from phase change are eliminated from the time evolution equations
\cite{Betts1973,Tripoli1981,Hauf1987}.
We also include the effects of the specific heats of both water
vapor and liquid water,
and consider an isentropic expansion factor ($\gamma$) that accounts 
for variations in the water composition of moist air.
Although the low Mach number formulation holds for any moist 
equation of state, for the purposes of numerical comparison with
compressible solutions we consider the special case where both 
dry air and water vapor are assumed to be ideal gases. 

While the larger time step allowed by the low Mach number formulation
can lead to greater computational efficiency than a purely compressible
formulation, it may also introduce 
larger errors in the dynamics of moist flows as investigated in \cite{CASTROmoist}.
In addition to the latent heat release accompanying phase changes, thermodynamic properties 
such as the specific heat of moist air, as well as thermodynamic variables
such as temperature, depend on the composition of the moist air,
thus change over the time step.  This motivates our use of invariant
variables as prognostic variables, namely total water content
and a specific enthalpy of moist air that
accounts for both sensible and latent heats.
In models where source terms related to phase transition appear explicitly 
in the evolution equations, they are typically first neglected or lagged 
in time and then corrected or estimated during a given time step; 
the use of invariant variables removes the need of accounting
for such terms.
Nevertheless, the diabatic contribution of the latent heat release
must appear in the source term for the low Mach divergence constraint
on the velocity field.  In practice the latter involves computing the 
rate of evaporation (or condensation).
Since no analytical expression exists for this rate,
one of the most common ways to estimate it deduces it from
the change in water vapor content necessary to ensure 
that there is no supersaturated water vapor 
at the end of the time step (cf. \cite{Soong1973}).
This variation is measured with respect to an initial estimate
of water vapor that does not necessarily respect the saturation
requirements either because it was initially advected without
accounting for phase changes 
(see, e.g., \cite{Klemp1978,Grabowski1990,Bryan2002,Satoh2003,ONeill2013})
or because it considers a lagged 
evaporation rate from the previous time step
(see, e.g., \cite{Grabowski2002}).
In our model, because water vapor is not used as a prognostic variable,
we cannot compute this variation of water vapor.
Instead we adopt a different approach, similar to \cite{LipHem82},
that estimates the evaporation rate based on the fact that
whenever a parcel is saturated, a Clausius--Clapeyron formula
relates the local values of water content to the thermodynamics
within the parcel.
The conservation equation for saturated water vapor then becomes a
time--varying constraint that guarantees thermodynamic equilibrium
from which the evaporation rate can be estimated.
The latter is thus computed as required during a 
time step using the current values of 
water content and thermodynamic variables,
diagnostically recovered from the invariant prognostic variables.

This paper is organized as follows.
We introduce the new low Mach number model for moist 
atmospheric flows in Section~\ref{sec:gov_eqs},
and describe the moist thermodynamics in Section~\ref{sec:moist_thermo}.
In Section~\ref{sec:numer} we discuss the numerical implementation.
Finally, in Section~\ref{sec:simul}, we present 
several numerical comparisons and discuss our findings.

\section{Low Mach number formulation}\label{sec:gov_eqs}
We begin by writing the fully compressible equations of motion
expressing conservation of mass, momentum, and enthalpy 
in a constant gravitational field
in which we neglect Coriolis forces and viscous terms:
\begin{eqnarray}
\frac{\partial \rho}{\partial t} + \Dx \cdotb \left( \rho \vel \right) 
&=& 0, \label{eqn:cont} \\
\frac{\partial \left( \rho \vel\right) }{\partial t} + 
\Dx \cdotb \left( \rho \vel \vel \right ) + \Dx p & = & -\rho g\ez, 
\label{eqn:momentum}\\
\frac{\partial ( \rho  \hhat ) }{\partial t} + 
\Dx \cdotb ( \rho \hhat \vel ) - \frac{D p}{D t}
& =  & 
\rho \Hhat,\label{eqn:enthalpy}
\end{eqnarray}
where the material derivative
is defined as
$D/D t = \partial /\partial t + \vel \cdotb \Dx$. 
Here $\rho$ is the total density, $\vel$ is the velocity,
and $\hhat$ is the specific enthalpy of moist air.
The pressure, $p$, is defined by an equation of state (EOS).
The term $\Hhat$ represents a source of heat (per unit mass and time)
to the system, such as thermal conduction or radiation.
We include gravitational acceleration given by
$\grav = -g \ez$, where $\ez$ is the unit vector in the vertical 
direction. 

We formulate the moist atmospheric processes as in \cite{Romps2008},
with the additional simplification that at any grid point all phases 
have the same velocity and temperature. 
We treat moist air as an ideal mixture of dry air, water vapor, 
and liquid water,  with the water phases in thermodynamic equilibrium,
so that only reversible processes are taken into account;
ice--phase microphysics, precipitation fallout, and subgrid--scale turbulence
are ignored.  Denoting by $q_a$, $q_v$, and $q_l$
the mass fraction of dry air, water vapor,
and liquid water, respectively, we note that $q_a+q_v+q_l = 1$ 
and write
\begin{eqnarray}
 \frac{\partial \left( \rho q_a \right)}{\partial t}  
+ \Dx \cdotb \left( \rho q_a \vel \right) & = & 0, \label{eqn:qa}\\
\frac{\partial \left( \rho q_v \right)}{\partial t}  
+ \Dx \cdotb \left( \rho q_v \vel \right) & =  & e_v, \label{eqn:qv}\\
\frac{\partial \left( \rho q_l \right)}{\partial t}  
+ \Dx \cdotb \left( \rho q_l \vel \right) & = & -e_v,\label{eqn:ql}
\end{eqnarray}
where $\rho$ is the total density.
Positive values of $e_v$ correspond to evaporation;
negative values correspond to condensation.
If we define the mass fraction of total water, $q_w =q_v +q_l$, 
then we can replace (\ref{eqn:qv})--(\ref{eqn:ql}) by
\begin{equation}\label{eqn:qw}
\frac{\partial \left( \rho q_w \right)}{\partial t}  
+ \Dx \cdotb \left( \rho q_w \vel \right) = 0.
\end{equation}
(We note that the system including both (\ref{eqn:cont}) and (\ref{eqn:qa})--(\ref{eqn:ql}) 
or (\ref{eqn:cont}), (\ref{eqn:qa}), and (\ref{eqn:qw}) is over--specified;
in practice (\ref{eqn:cont}) need not be solved separately.)

We define the internal energy of
moist air, $\ehat,$ as in \cite{Romps2008},
\begin{equation}\label{eqn:ehat}
\ehat = \cvm \left(T - \Ttrip  \right) + q_v \Eov,
\end{equation}
where the constant--volume specific heat of moist air is
given by
$\cvm = q_a \cva + q_v \cvv + q_l \cvl,$
with the specific heats at constant volume:
$\cva$, $\cvv$, and $\cvl$, for 
dry air, water vapor, and liquid water, respectively.
Here $\Ttrip$ is the triple--point temperature
and $\Eov$ is the specific internal energy of water vapor
at the triple point,
while the enthalpy of moist air is given by
\begin{equation}\label{eqn:hhat}
\hhat = \ehat + \frac{p}{\rho}.
\end{equation}
Note that an evolution equation for the internal energy
\begin{equation}\label{eqn:int_energy}
\frac{\partial \left( \rho  \ehat \right) }{\partial t} + 
\Dx \cdotb \left( \rho \ehat \vel \right) 
+ p \left(\Dx \cdotb \vel\right) = 
\rho \Hhat,
\end{equation}
can be used instead of (\ref{eqn:enthalpy}).
In either case, there are no source terms related to phase change
in (\ref{eqn:enthalpy}) or (\ref{eqn:int_energy}),
as observed in \cite{Ooyama1990} and \cite{CASTROmoist}.
Finally, to close the system
(\ref{eqn:cont})--(\ref{eqn:qa}) with (\ref{eqn:qw}),
we consider a general equation of state for moist air written as
$p=p(\rho,T,q_a,q_v,q_l)$.  (Because 
$q_a+q_v+q_l = 1$ we could remove one of these three arguments from 
the equation of state; for clarity of exposition, however, we leave 
all three.)

In the low Mach number model, we write the pressure, $p(\x,t)$,
with $\x=(x,y,z)$, as the sum of a base state pressure,
$\po(z,t)$, and a perturbational, or dynamic, pressure,
$p'(\x,t)$, such that 
$|p'|/\po = O(M^2)$.
In contrast to anelastic models, 
the perturbations in density and temperature themselves do not need to 
be small for the equations to remain valid;
as long as those perturbations do not result in the flow violating the
assumption of a low Mach number, hence small $|p'|/\po$, the equations
remain valid. 
The base state is assumed to be in hydrostatic equilibrium,
i.e., $\Dx\po=-\rhozero g \ez$, where $\rhozero=\rhozero(z,t)$ is the
base state density.
The fully compressible momentum equation (\ref{eqn:momentum})
could be rewritten as
\begin{equation*}\label{eqn:momentumLM_orig}
\frac{\partial \left( \rho \vel\right) }{\partial t} + 
\Dx \cdotb \left( \rho \vel \vel \right ) + \Dx p' = 
-(\rho-\rhozero)g\ez, 
\end{equation*}
with no approximation.  The low Mach number momentum equation
has an additional contribution to the buoyancy term,
\begin{equation}\label{eqn:momentumLM}
\frac{\partial \left( \rho \vel\right) }{\partial t} + 
\Dx \cdotb \left( \rho \vel \vel \right ) + \Dx p' = 
-(\rho-\rhozero)g\ez - 
\left( \frac{\rhozero}{\rho} 
\left.\frac{\partial \rho}{\partial \po}\right \vert_s p' \right)g\ez,
\end{equation}
as introduced by \cite{Klein2012}.  Here the derivative of $\rho$
with respect to $p_0$ is taken at constant entropy, $s,$ 
and is based on the low Mach number form of the equation of state, i.e. 
$p_0(z,t) = p(\rho,T,q_a,q_v,q_l)$.  As pointed out in \cite{VLBWZ:2013},
this form of the momentum equation is analytically equivalent to the 
momentum equation in the pseudo--incompressible equation set, i.e., 
\begin{equation*}\label{eqn:momentum_pseudo}
\frac{\partial \vel}{\partial t} + 
\vel \cdotb \Dx \vel + \cpa \theta \Dx \pi^\prime = 
\frac{\theta - \theta_0}{\theta_0} g \ez
\end{equation*}
with Exner pressure, $\pi = \left( p /  p_{00} \right)^{R_a/\cpa}$
($p_{00}$ is a reference pressure,
while $R_a$ and $\cpa$ stand for the specific gas constant
and constant--pressure specific heat of dry air, respectively),
potential temperature, $\theta = T / \pi,$ 
and the base state potential temperature, $\theta_0$,
if we define
a linearized relationship between the perturbational pressures,
$p^\prime$ and $\pi^\prime.$ 

To complete the low Mach number model for moist atmospheric flows
we first replace $p$ by $p_0$ in the equation of state
and we then 
differentiate $p_0 = p(\rho,T,q_a,q_v,q_l)$ along particle paths.
Following the derivation in \cite{ABNZ:III},
with details as given in Appendix \ref{app:LM_constraint}, 
we derive the following divergence constraint 
for the low Mach number model:
\begin{equation}\label{eqn:divu_alpha}
\Dx \cdotb \vel
+ \alpha \frac{D \po}{D t} = S,
\end{equation}
with 
\begin{equation}\label{eqn:S}
\alpha = \frac{1}{\Gamma_1 p_0},
\qquad
S = 
\left[
\frac{1}{\rho p_{\rho}} \left(p_{q_v} - p_{q_l} \right)
- \sigma \left(\hhat_{q_v} - \hhat_{q_l} \right) 
\right]
\frac{e_v}{\rho}
+ \sigma \Hhat,
\end{equation}
where
$p_{\rho} = \partial p/\partial \rho \vert_{T,q_i}$,
$p_{T} = \partial p/\partial T \vert_{\rho,q_i}$,
$p_{q_i} =  \partial p/\partial q_i\vert_{\rho,T,(q_j,j\neq i)}$,
$\hhat_p = \partial \hhat/\partial p \vert_{T,q_i}$,
$\hhat_{q_i} = \partial \hhat / \partial q_i \vert_{T,p,(q_j,j\neq i)}$,
$\Gamma_1 = {\partial (\log p)}/{\partial (\log \rho)}\vert_s$, and
$\sigma = p_{T}/\left(\rho \cpm p_{\rho} \right)$,
where $\cpm = \partial \hhat/\partial T \vert_{p,q_i}$
is the specific heat of moist air at constant pressure.
Here, as in (\ref{eqn:momentumLM}), $\vert_s$ refers to the derivative at
constant entropy.
Notice that both $\alpha$ and $S$ in the divergence constraint
(\ref{eqn:divu_alpha}) depend on the water composition of
moist air, $q_v$ and $q_l$.
Most importantly the constraint on the divergence of the velocity field retains
compressibility effects from stratification as well as latent heat release 
and compositional changes.

Summarizing the low Mach number equation set
for moist atmospheric flows in the form we will use, we have
\begin{eqnarray}
\frac{\partial (\rho q_a)}{\partial t}  &=& - \Dx \cdotb (\rho q_a \vel) 
\label{eq:rhoqaupd} \\
\frac{\partial (\rho q_w)}{\partial t}  &=& - \Dx \cdotb (\rho q_w \vel) 
\label{eq:rhoqwupd} \\
\frac{\partial (\rho \hhat)}{\partial t}  &=& - \Dx \cdotb (\rho \hhat \vel) 
+ \frac{Dp_0}{Dt} + \rho \Hhat, \label{eq:rhohhatupd} \\
\frac{\partial\vel}{\partial t} &=& -\vel \cdotb \Dx \vel
 -\frac{1}{\rho} \Dx p' - \frac{(\rho - \rhozero)}{\rho} g \ez
- \left( \frac{\rhozero}{\rho^2} 
\left.\frac{\partial \rho}{\partial \po}\right \vert_s p' \right)g\ez, 
\label{eq:velupd} \\
 \Dx \cdotb \vel & = & -
\alpha \frac{D \po}{D t} + S, \label{eq:constr}
\end{eqnarray}
where the total mass density is defined as
\begin{equation}
\rho =  \rho q_a + \rho q_w.
\end{equation}
Contrary to the original compressible equation set
(\ref{eqn:cont})--(\ref{eqn:enthalpy}) together
with (\ref{eqn:qa}) and (\ref{eqn:qw}) and a given equation
of state,
the total pressure field, $p$, is now decoupled 
from the density and the other state variables
in the enthalpy and momentum equations 
(\ref{eq:rhohhatupd}) and (\ref{eq:velupd}) 
(to be compared with (\ref{eqn:enthalpy}) and (\ref{eqn:momentum})).
Instead of having $\Dx p$ in the momentum equation (\ref{eq:velupd}),
we now have $\Dx p'$, where $p'$ is a perturbational pressure on the
background pressure $\po$, that is no longer related to the other
state variables
through the equation of state.
That is, pressure is no longer advanced in time through system 
(\ref{eq:rhoqaupd})--(\ref{eq:constr}),
and hence acoustic modes are filtered out from the original
compressible governing equations.
In practice, $p'$ is computed in such a way that 
the divergence constraint (\ref{eq:constr}) is satisfied at a given time,
analogous to what is done for incompressible flows.
While the equation of state might at first glance appear to be missing
from the low Mach number equation set, it is in fact encapsulated in
the constraint, (\ref{eq:constr}), which was derived by 
differentiation of the equation of state along particle paths.  
This differs substantially from the anelastic approximation which results
from 
replacing the density 
by the background density in the continuity equation, which thus
contains no information about the equation of state.

An underlying assumption in the low Mach number approximation is that the
pressure remains close to the background pressure.
Heat release and large--scale convective motions
in a convectively unstable background
can both cause the background state to evolve in time. 
As discussed in \cite{almgren:2000} and demonstrated numerically in
\cite{ABRZ:II} for an externally specified heating profile,
if the base state does not evolve in response to heating,
the low Mach number method quickly becomes invalid.  
For the small--scale motions of interest here, the base state can 
effectively be viewed as independent of time; however, for the sake
of completeness we retain the full time dependence of the base state
in the development of the methodology.

\section{Moist thermodynamics}\label{sec:moist_thermo}
Phase changes and thus variations in 
water composition of moist air
are introduced in the flow dynamics 
through the divergence constraint (\ref{eq:constr}),
specifically through $\alpha$ and $S$.
These parameters are evaluated at a given time 
accounting for the current water composition 
in terms of liquid and vapor, and thus accounting for
phase transitions and the current saturation requirements,
given the prognostic state variables.
To define $\alpha$ and $S$ according to (\ref{eqn:S}) 
in the divergence constraint (\ref{eq:constr}),
we here consider dry air and water vapor to be ideal gases
(see, e.g., \cite{Ooyama1990,Satoh2003,Klemp2007}), and note
that while the low Mach number formulation allows a more general
equation of state, this is a standard assumption in atmospheric modeling.

The partial pressures of dry air and water vapor are then given by
$p_a = \rho q_a R_a T$ and 
$p_v = \rho q_v R_v T,$
where $R_a = R/M_a$ and $R_v = R/M_v$
are the specific gas constants for dry air and water vapor, respectively,
with $R$ the universal gas constant for ideal gases,
and $M_a$ and $M_v$ the molar masses of dry air and water, respectively. 
The sum of the partial pressures defines the total pressure of a parcel,
\begin{equation}\label{eqn:pres}
p = p_a + p_v = \rho R_m T,
\end{equation}
where the specific gas constant of moist air is defined as
\begin{equation*}\label{eqn:R_m}
R_m = q_aR_a + q_vR_v = \left(\frac{q_a}{M_a} + \frac{q_v}{M_v}\right) R.
\end{equation*}
Additionally, the specific heat capacities at constant
pressure can be defined as
\begin{equation*}\label{eqn:cpm}
\cpa = \cva + R_a, \quad
\cpv = \cvv + R_v, \quad
\cpm = \cvm + R_m,
\end{equation*}
for dry air, water vapor, and moist air, respectively.
A common approximation in cloud models 
is to neglect the specific heats of water vapor and liquid water
(see, e.g., \cite{Bryan2002} for a study and discussion on this topic).
This was also assumed in the moist pseudo--incompressible model
of \cite{ONeill2013}.
Here we consider specific heats for all three phases.

With this choice for the equation of state for moist air
(eq.~(\ref{eqn:pres})),
we have 
\begin{equation}\label{eqn:alphapres}
\alpha = \frac{1}{\gamma_m \po},
\qquad
S = 
\left[
\frac{1}{(\epsilon q_a +  q_v)} 
- \frac{L_e}{\cpm T}
\right]
\frac{e_v}{\rho}
+ \left[ \frac{1}{\cpm T}\right] \Hhat,
\end{equation}
where $\gamma_m= \cpm / \cvm$ is the 
isentropic expansion factor of moist air, 
$\epsilon = R_a/R_v=M_v/M_a$,
and the latent heat of vaporization, $L_e$,
is defined as
\begin{equation}\label{eqn:Le}
L_e = \Eov + R_v T + (\cvv - \cvl)(T - \Ttrip).
\end{equation}
Note that both $\alpha$ and $S$ depend
on the composition of the moist mixture even 
when there is no phase transition, that is, 
when $e_v =0$. We can now replace
$\Gamma_1$ as used in (\ref{eqn:divu_alpha})--(\ref{eqn:S})
by $\gamma_m$ for moist atmospheric flows.

The saturation vapor pressure 
with respect to liquid water, $\pvstar$, is defined 
by the following Clausius--Clapeyron relation:
\begin{equation}\label{eqn:pvstar}
\pvstar(T) = \ptrip \left(\frac{T}{\Ttrip}\right)^{\aav}
\exp \left [ \bbv 
\left( \frac{1}{\Ttrip}-\frac{1}{T} \right) \right ],
\end{equation}
with constants $\aav$ and $\bbv$ given, for instance, by 
\begin{equation}\label{eqn:constant_pvstar}
\aav = \frac{\cpv-\cvl}{R_v}, \qquad 
\bbv = \frac{\Eov -(\cvv-\cvl)\Ttrip}{R_v},
\end{equation}
as in \cite{Romps2008}.
The saturated mass fraction of water vapor,
$\qvstar$, can be then computed from the EOS,
given in this case by
\begin{equation}\label{eqn:qvstar}
\qvstar(\rho,T) = \frac{\pvstar}{\rho R_v T},
\end{equation}
or equivalently by
\begin{equation}\label{eqn:qvstar_2}
\qvstar(q_a,p,T) = \frac{\epsilon q_a \pvstar}{p - \pvstar}.
\end{equation}
Following \cite{Ooyama1990,Satoh2003}, we assume 
that air parcels cannot be supersaturated,
and thus $q_v$ cannot exceed its saturated value, $\qvstar$.
The water mass fractions,
$q_v$ and $q_l$, as well as the temperature
$T$, which are not explicitly computed in 
(\ref{eq:rhoqaupd})--(\ref{eq:constr}),
are obtained by solving the following
nonlinear system of equations:
\begin{equation}\label{eqn:solveT}
\left.
\begin{array}{l}
\hhat =
\cpm(q_a,q_v,q_l)  \left(T - \Ttrip  \right) + 
R_m(q_a,q_v) \Ttrip + 
q_v \Eov, \\[1.5ex]
q_v = \min\left[\qvstar(\rho,T),q_w\right],  \\[1.5ex]
q_l = q_w - q_v,
\end{array}
\right\}
\end{equation}
which satisfies the Clausius--Clapeyron relation
and the saturation requirements, given $(\rho,\hhat,q_a,q_w)$.

No approximation has been introduced so far to evaluate
the water composition and therefore the thermodynamic
properties of the moist fluid.
We now need to estimate the evaporation rate, $e_v$,
that is required to quantify the latent heat release
in the divergence constraint.
Considering that there is no analytical expression
for such a rate and that we cannot derive it from an approximate
value of vapor water as previously discussed in the Introduction,
we introduce the following approach.
Taking into account that 
the evaporation rate
$e_v$ is different from zero
only when a change of phase is taking place, that is, when 
$q_v = \qvstar$,
we can rewrite the conservation equation for $q_v$ (eq.~(\ref{eqn:qv})) as
\begin{equation}\label{eqn:ev_qvstar}
e_v =
\frac{\partial \left( \rho \qvstar \right)}{\partial t}  
+ \Dx \cdotb \left( \rho \qvstar \vel \right) = 
\rho \frac{D \qvstar}{D t}.
\end{equation}
We show in Appendix \ref{app:evp_rate}
that if $q_v = \qvstar$, then
\begin{equation}\label{eqn:ev_hhat_rho_a}
e_v = \Aev \Dx \cdotb \vel + 
\Bev \frac{D \po}{D t} + \Cev \rho \Hhat,
\end{equation}
where 
parameters
$\Aev$, $\Bev$, and $\Cev$
are in general functions of $(\rho,T,q_a,q_v,q_l)$.
Otherwise, $e_v=0$ whenever $q_v < \qvstar$.
In \cite{LipHem82}, the term 
$\Dx \cdotb ( \rho \qvstar \vel)$ in (\ref{eqn:ev_qvstar})
is replaced with 
$\Dx \cdotb ( \rho q_{va} \vel )$,
where $( \rho q_{va} )$ stands for the 
water vapor content advected without considering
the evaporation rate in (\ref{eqn:qv}):
$\partial ( \rho q_{va}) / \partial t + 
\Dx \cdotb \left( \rho q_{va} \vel \right) = 0$.

Finally, using (\ref{eqn:ev_qvstar})
to estimate $e_v$ 
in $S$ (eq. (\ref{eqn:alphapres}))
involves having a low Mach divergence constraint 
where the source term depends on 
the pressure and the velocity field, i.e.,
\begin{equation}\label{eqn:divu_alpha_dependency}
\Dx \cdotb \vel
+ \alpha \frac{D \po}{D t} = 
S\left(\Dx \cdotb \vel , \frac{D \po}{D t} \right).
\end{equation}
A simpler expression can be obtained
by rearranging the terms in (\ref{eqn:divu_alpha_dependency}),
as shown in Appendix \ref{app:mod_cons},
which leads to
a modified divergence constraint:
\begin{equation}\label{eqn:divu_ev}
\Dx \cdotb \vel
+ \talpha \frac{D \po}{D t} = \tS,
\end{equation}
similar to the general low Mach divergence constraint 
(\ref{eqn:divu_alpha}),
with 
\begin{equation}\label{eqn:alphatilde}
\talpha = \frac{1}{\tgamma \po},
\qquad \tS = \tsigma \Hhat,
\end{equation}
both depending on 
$(\rho,T,q_a,q_v,q_l)$.
In particular
if no heat source is considered, 
we have that $\tS = 0$ and 
all the information related to phase change 
is included in $\talpha$.
Now  
$\Gamma_1$ 
in the general low Mach constraint
(\ref{eqn:divu_alpha})--(\ref{eqn:S})
is given by 
a modified isentropic expansion factor of moist air:
$\tgamma$ 
(eq. (\ref{eqn:gammatilde1}) in Appendix \ref{app:mod_cons}).
Both divergence constraints, (\ref{eqn:divu_alpha_dependency})
and (\ref{eqn:divu_ev}), are analytically equivalent.

\section{Numerical methodology}\label{sec:numer}
In order to solve the low Mach number equation set 
(\ref{eq:rhoqaupd})--(\ref{eq:constr}) we begin with the MAESTRO code,
which was originally designed to simulate
low Mach number stratified, reacting flows in astrophysical settings
\cite{ABRZ:I,ABRZ:II,ABNZ:III,multilevel}. 

We recall that for the moist equation of state considered here
we replace $\Gamma_1$ in the original
MAESTRO notation by $\gamma_m$. Since $\gamma_m$ in general varies in 
space and time, the solution procedure in the original
MAESTRO algorithm replaces 
$\gamma_m(\x)$ by  $\gammambar(z)$,
that is, 
$\alpha (z) = 1/(\gammambar \po)$
in the divergence constraint
(\ref{eqn:divu_alpha_dependency})
where 
$\gammambar$
is the lateral average
of 
$\gamma_m$,
\begin{equation}\label{eq:lateral_average}
\gammambar(z) = \frac{1}{A(\Omega_H)} \int_{\Omega_H} \gamma_m(\x) \; d\Omega,
\end{equation}
where ${\mathrm A}(\Omega_H) = \int_{\Omega_H} d\Omega,$ $\Omega_H$ is
a region at constant height for the plane--parallel atmosphere,
and $d\Omega$ represents an area measure.
The same follows for 
$\tgamma$ in the modified divergence constraint
(\ref{eqn:divu_ev}):
$\talpha (z)= 1/(\tgammambar \po)$.
The introduction of an averaged 
$\gamma_m$ (or $\tgamma$)
allows us to rewrite 
the constraint 
(\ref{eqn:divu_alpha_dependency})
(or (\ref{eqn:divu_ev})) as
\begin{equation}\label{constraint_eq}
\Dx \cdotb (\beta_0 \vel) = \beta_0 \left (S - \alpha \frac{\partial p_0}{\partial t}\right),  
\end{equation}
as shown in \cite{ABRZ:I} (Appendix B), 
with
\begin{equation*}\label{eq:beta_0}
\beta_0(z,t) = \beta(0,t) \exp \left ({\int_0^z 
\alpha (z)
               \frac{\partial p_0}{\partial z^\prime} \, dz^\prime} \right ),
\end{equation*}
(with $\tS$ and $\talpha$ instead of $S$ and $\alpha$
if the modified divergence (\ref{eqn:divu_ev}) is considered).
Moreover,
the momentum equation (\ref{eq:velupd}), including the 
correction term introduced by \cite{Klein2012},
can be equivalently recast as
\begin{equation}\label{eqn:momentumLM_beta}
\frac{\partial\vel}{\partial t} = -\vel \cdotb \Dx \vel
 -\frac{\beta_0}{\rho} \Dx \left(\frac{p'}{\beta_0} \right) 
- \frac{(\rho - \rhozero)}{\rho} g \ez,
\end{equation}
as derived in \cite{VLBWZ:2013}, and discussed in \cite{ABNZ_energy}.

MAESTRO thus solves the low Mach equation set 
(\ref{eq:rhoqaupd})--(\ref{eq:rhohhatupd})
with the momentum equation (\ref{eqn:momentumLM_beta})
and the divergence constraint (\ref{constraint_eq}).
A predictor--corrector formalism is implemented
to solve the flow dynamics, as detailed in \cite{multilevel}.
In the predictor step
an estimate of the expansion of the base state is first computed,
and then an estimate of the flow variables at the new time level.  
In the corrector step
results of the previous state update are used to compute a new base
state update as well as the full state update.

Since we are considering the time evolution of total water 
(\ref{eq:rhoqwupd})
and the definition of enthalpy of moist air (\ref{eqn:hhat})
involves a conservation equation (\ref{eq:rhohhatupd})
without source terms related to phase change,
all the information related to variations in the moist
composition and latent heat release 
is contained in the divergence constraint
(\ref{constraint_eq}).  
Here
$q_v$ and $q_l$, as well
as the local temperature $T$, must be computed point--wise 
in order to define $\alpha$, $S$, and $\beta_0$
(or $\talpha$, $\tS$, and $\beta_0$).
Point--wise values of $(q_v,q_l,T)$ are thus diagnostically recovered 
by solving the nonlinear system (\ref{eqn:solveT}),
given the values of $(\rho,\hhat,q_a,q_w)$ at a given time; 
we use the Newton solver described in \cite{CASTROmoist}.

We consider two approaches for handling phase transitions
depending on which divergence constraint is used
to define (\ref{constraint_eq})
in the numerical implementation:
(\ref{eqn:divu_alpha_dependency})
or (\ref{eqn:divu_ev}).
In the first case, 
which uses (\ref{eqn:divu_alpha_dependency}),
the evaporation rate 
is evaluated according to (\ref{eqn:ev_hhat_rho_a})
and introduced in the source term $S$ of the constraint.
As previously remarked, there is a dependence
of $S$ on the velocity field and the base state pressure.
Consequently, approximate or lagged values of $\vel$ and $\po$ are used to
estimate $e_v$ during the prediction step, which
is later recomputed with the updated values of 
$\vel$ and $\po$ during the correction step.
Notice, however, that during both steps 
the moist composition and hence phase transitions are
diagnostically recovered based on the current values
of $(\rho,\hhat,q_a,q_w)$.
The 
second approach
consists of considering
the modified divergence constraint (\ref{eqn:divu_ev}),
in which case there is no need to estimate the evaporation rate 
(\ref{eqn:ev_hhat_rho_a}).
Parameters $\talpha$, $\tS$, and $\beta_0$ in 
(\ref{constraint_eq}) are computed with
$(\rho,T,q_a,q_v,q_l)$ given the current values
of $(\rho,\hhat,q_a,q_w)$ throughout the predictor--corrector
scheme for the flow dynamics.

Notice that considering the 
lateral average of $\gamma_m(\x)$ through 
(\ref{eq:lateral_average}) is the only approximation
introduced in the present model in terms of thermodynamic
properties of the moist fluid.
The replacement of $\Gamma_1(\x,t)$ by its 
lateral average 
$\gammabar (z,t)$ was demonstrated
in \cite{ABNZ:III} to have little effect on the astrophysical
flows studied there.
In the case of moist atmospheric flows,
$\Gamma_1$ is given by $\gamma_m$
which varies according to the local
moist air composition at a given time and position.
If the modified divergence constraint (\ref{eqn:divu_ev})
is considered, 
$\Gamma_1$ is given by $\tgamma$ which varies 
not only according to the moist air composition 
but also the local values of $(\rho,T)$.
We can thus 
consider the effects of localized variations in 
$\Gamma_1$ 
following \cite{ABNZ:III},
by writing $\Gamma_1(\x,t) = \gammabar(z,t) + \delta \Gamma_1(\x,t)$,
and hence,
\begin{equation*}
\Dx \cdotb \vel  
 = - \frac{1}{(\gammabar + \delta \Gamma_1) \po} 
  \frac{D\po}{Dt} +
S,
\end{equation*}
instead of (\ref{eq:constr}).
Assuming that $\delta \Gamma_1 \ll \gammabar$, we then have
\begin{equation}\label{eq:gammafull}
\Dx \cdotb \vel = 
- \frac{1}{\gammabar \po} \frac{D\po}{Dt}
+ S  +
\frac{\delta \Gamma_1}{\gammabar^2 \po} 
\frac{D\po}{Dt}, 
\end{equation}
which leads to 
\begin{equation}\label{constraint_eq_deltagamma1}
\Dx \cdotb (\beta_0 \vel) = 
\beta_0 \left (S - \alpha \frac{\partial p_0}{\partial t}
+
\frac{\delta \Gamma_1}{\gammabar^2 \po} 
\frac{D\po}{Dt}
\right).
\end{equation}
We will refer to the $\delta \Gamma_1$--correction whenever
(\ref{constraint_eq_deltagamma1}) is considered instead of
(\ref{constraint_eq}).
Notice that 
$D\po/D t = \partial \po/\partial t + \vel \cdotb \Dx\po$,
and therefore, to solve (\ref{constraint_eq_deltagamma1})
a lagged $\vel$ is used in evaluating the right--hand side,
as described in \cite{ABNZ:III}.

\section{Numerical simulations}\label{sec:simul}
To validate the low Mach number method we compare simulations
using the low Mach number method to simulations using a
fully compressible approach.
The first problem we consider is a benchmark problem 
presented in \cite{Bryan2002} for moist flows
in an isentropically stratified background.
We investigate both the first and second approach to implementing phase 
transitions, and find very good agreement between the
low Mach number and compressible simulations.
We then consider a problem described in \cite{GrabowskiClark1991} 
for non--isentropic background states and both saturated and only 
partially saturated media, which was also 
studied in \cite{CASTROmoist}.
Finally, we show a comparison of three dimensional simulations.

\subsection{Isentropic background state}

Two--dimensional simulations of a benchmark test case are investigated in 
\cite{Bryan2002}.  
The computational domain is $10\,$km high and $20\,$km wide;
the background atmosphere is isentropically stratified,
at a uniform wet equivalent potential temperature, $\theta_{e0}=320\,$K,
and is saturated, that is, $q_v = \qvstar$ and $q_l >0$
everywhere in the domain.  A warm perturbation in potential temperature
is introduced in the domain, which leads to a rising thermal.
Here we use the configuration and parameters as given in \cite{CASTROmoist}.
The normal velocity is set equal to zero on the side and bottom
boundaries; at the top boundary the normal velocity is set equal to the 
velocity corresponding to the base state evolution.  (See \cite{ABRZ:II}
for further detail about the base state evolution.)
Homogeneous Neumann boundary conditions are used in solving for the
perturbational pressure. 
For the thermodynamic variables, we impose
homogeneous Neumann boundary conditions on the horizontal sides;
the background state is reconstructed by extrapolation at 
vertical boundaries 
and the full state is set equal to the base one there,
in order to determine the corresponding fluxes.
The time step for the low Mach number computations 
is dictated by the advective CFL number
which is based on the velocity but not the sound speed; 
we set this to $0.9$.

In the first approach to account for phase transitions
the divergence constraint is given by 
(\ref{eqn:divu_alpha_dependency})
with $\alpha$ and $S$ defined by (\ref{eqn:alphapres}),
and the evaporation rate  (\ref{eqn:ev_hhat_rho_a}).
For this particular problem,
$\Hhat=0$ in $S$, and hence it is not necessary 
to compute $\Cev$ in (\ref{eqn:ev_hhat_rho_a}).
For a $256\times 128$ grid
of similar spatial resolution to that considered 
in \cite{Bryan2002},
the maximum and minimum values for 
the perturbational wet equivalent potential temperature
($\theta'_e = \theta_e - \theta_{e0}$) are 
$4.05402\,$K and $-0.28931\,$K, respectively,
compared with the original 
$4.09521\,$K and $-0.305695\,$K in \cite{Bryan2002}.
Additionally, our computation yields 
$15.8199\,$m s$^{-1}$ and $-9.45586\,$m s$^{-1}$,
for the maximum and minimum vertical velocities, respectively,
to be compared with  $15.7130\,$m s$^{-1}$ and $-9.92698\,$m s$^{-1}$
in \cite{Bryan2002}.  The low Mach number code takes roughly
a factor of six less computational time than the compressible
simulation at the same resolution run with the 
code described in \cite{CASTROmoist} at an acoustic CFL number
of $0.9$.

The second approach does not explicitly estimate $e_v$,
but rather considers a
modified divergence constraint (eq.~(\ref{eqn:divu_ev}))
with $\talpha$ and $\tS$ defined by (\ref{eqn:alphatilde}). 
Since for this particular problem, $\Hhat=0$, we have $\tS=0$.
For a $256\times 128$ grid, the maximum and minimum values for 
the perturbational wet equivalent potential temperature $\theta'_e$ 
are now given by $4.12911\,$K and $-0.30190\,$K, respectively.
The time steps taken with the second approach are 
slightly larger than with the first approach, and
the overall computational time is roughly $10$ to $15\,$\% 
less with this approach than with the first approach.
\begin{figure}[!ht]
\centering
\includegraphics[width=0.49\textwidth]{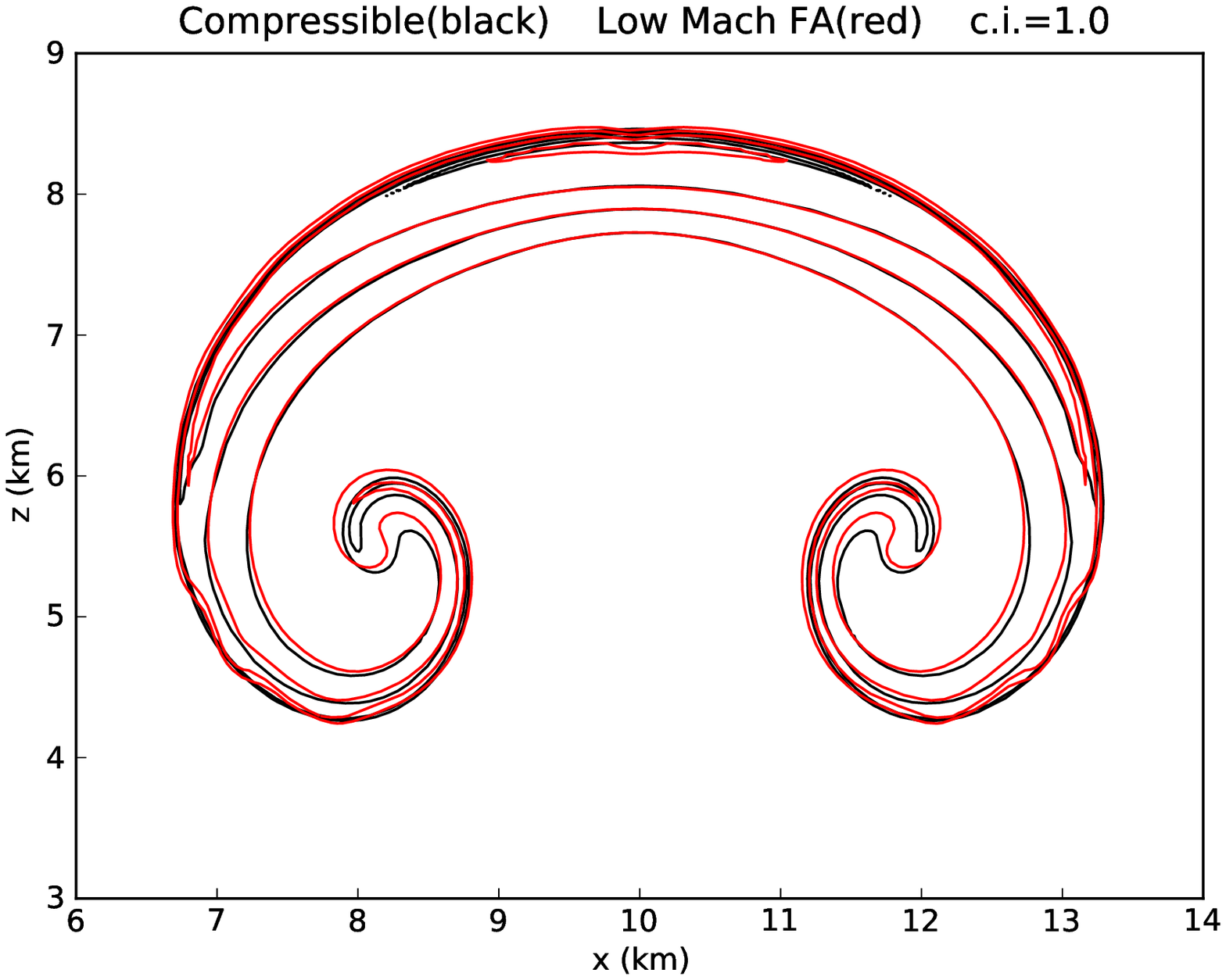}
\includegraphics[width=0.49\textwidth]{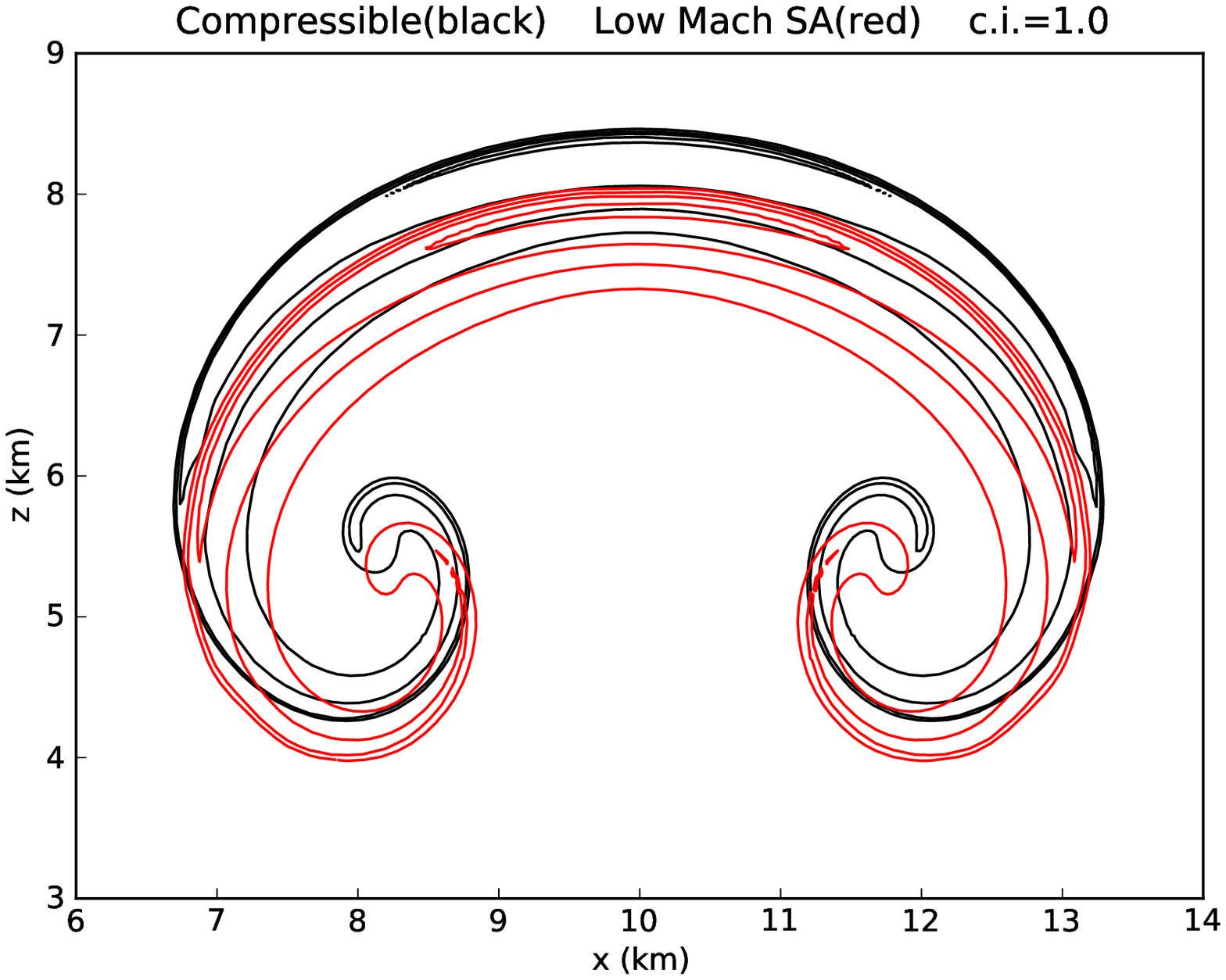}
\includegraphics[width=0.49\textwidth]{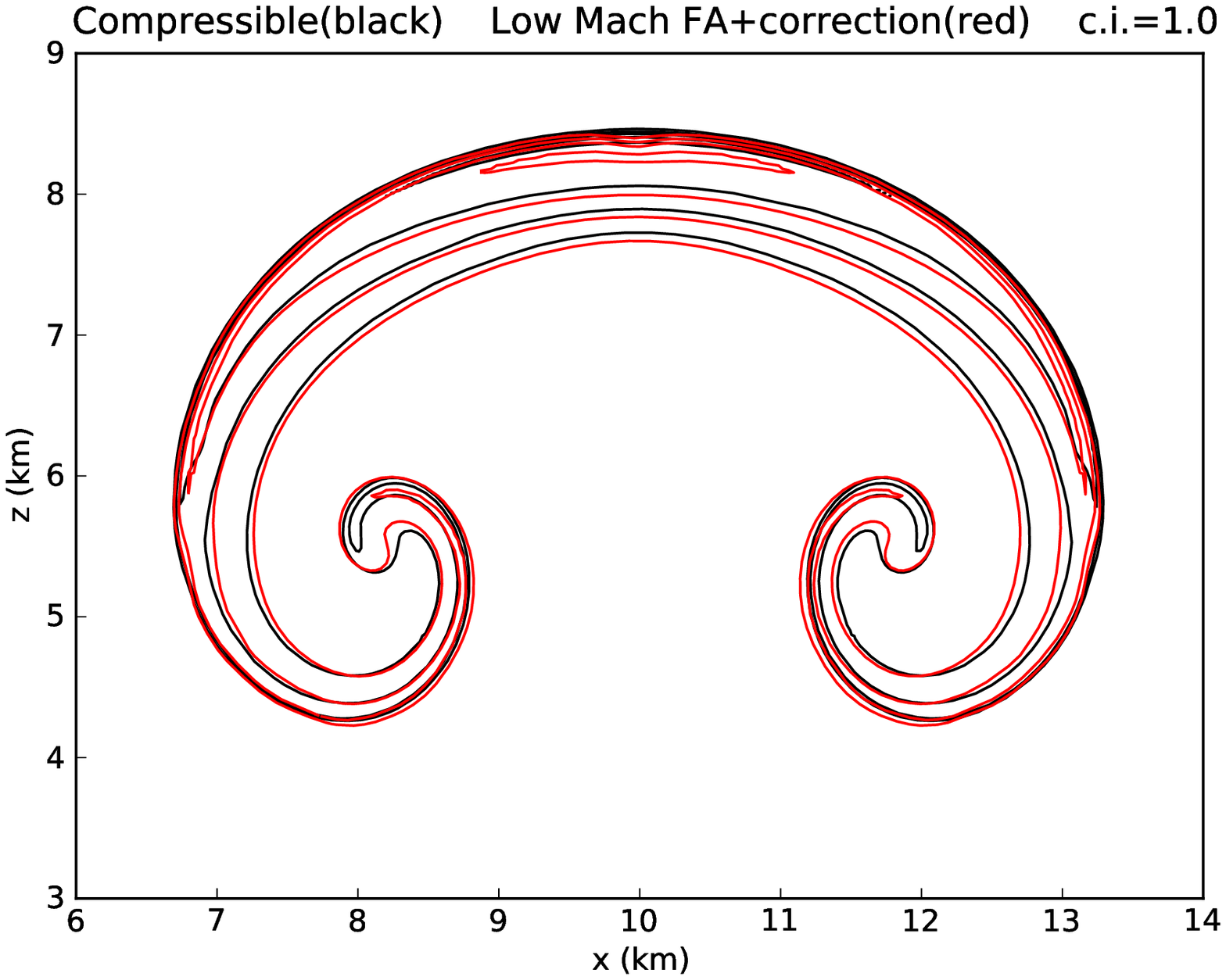}
\includegraphics[width=0.49\textwidth]{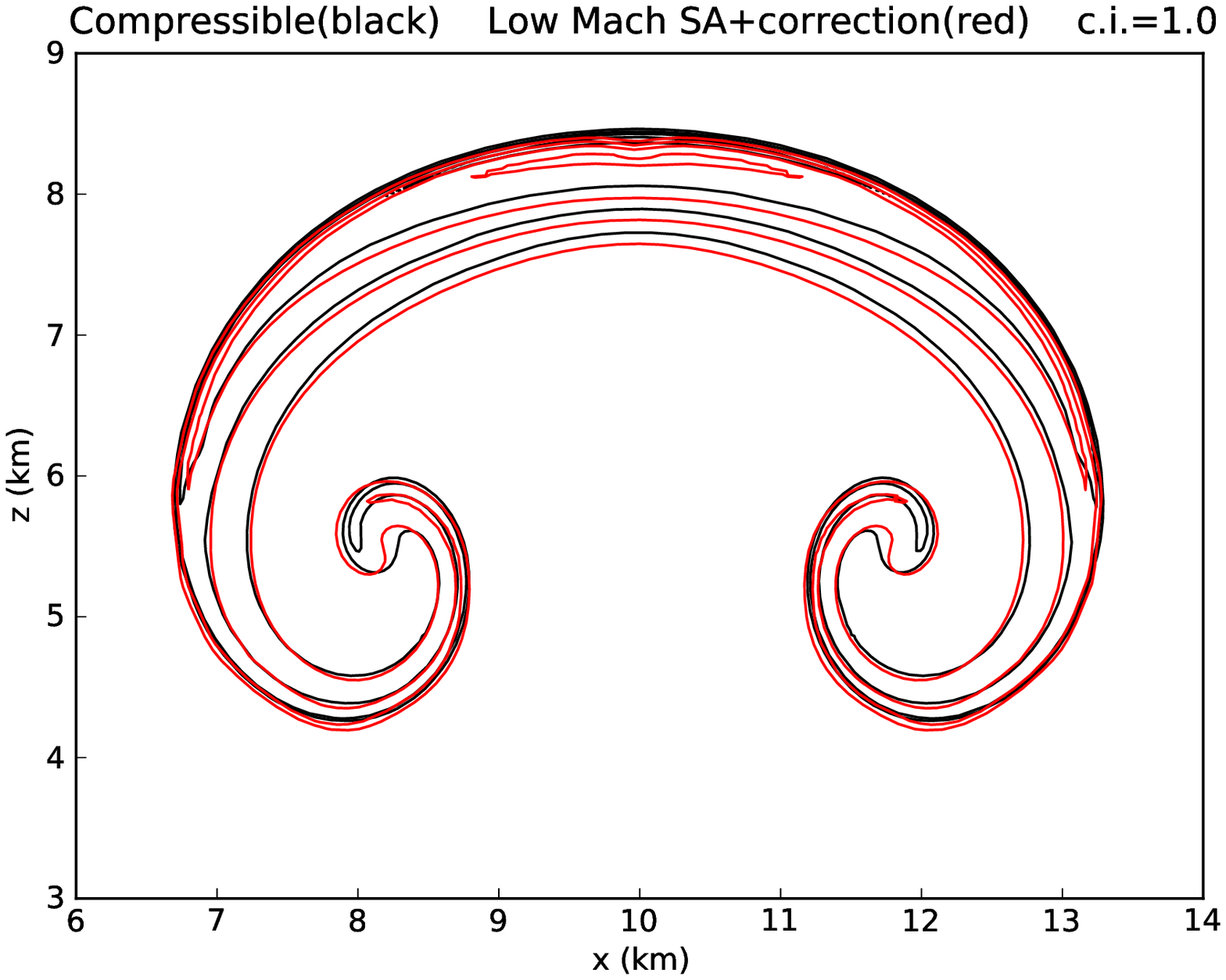}
\caption{Comparison with the compressible solution for the moist thermal simulations
at $1000\,$s.  
Perturbational potential temperature is shown with contours every $1\,$K:
first (left) and second approach (right)
The simulations in the bottom row include the $\delta \Gamma_1$--correction.
}
\label{fig:compressible_comp}
\end{figure}

Figure~\ref{fig:compressible_comp} shows results from computations
of the problem described above;  in this figure $\theta'_e$ from the low
Mach number simulations is overlaid on the reference compressible solution,
which was computed with the fully coupled compressible solver 
described in \cite{CASTROmoist} and validated against results from
\cite{Bryan2002}.  
The low Mach number simulations were carried out on a uniform grid 
of $512\times 256$;  the compressible solution was performed on a 
finer grid of $1024\times 512$ in order to get a more accurate 
reference solution.   In the top row of Figure~\ref{fig:compressible_comp} 
we see that the low Mach number computation using the
first approach agrees well with the compressible solution. The
computation with the second approach, however, shows a significant
height difference of the top of the thermal.
In the bottom row of Figure~\ref{fig:compressible_comp} we
see that  accounting for the local variation of $\gamma_m(\x)$
and $\tgamma (\x)$ through the the $\delta \Gamma_1$--correction described 
in \S\ref{sec:numer} improves the fidelity of both approaches.
With the first approach we see the improvement mostly in the tips;
with the second approach the height of the thermal is noticeably improved.
The fact that including the $\delta \Gamma_1$--correction 
impacts the second approach more dramatically than the first is consistent
with the observation that replacing $\gamma_m(\x)$ by $\gammambar$ 
in the first approach meant neglecting local variations in $\gamma_m(\x)$  
from $-8\times 10^{-5}$ to $6.5\times 10^{-4}$, while replacing 
$\tgamma  (\x)$ by $\tgammambar$ 
in the second approach meant neglecting local variations in $\tgamma  (\x)$ 
at least an order of magnitude larger, i.e. from $-6.7\times 10^{-3}$ to $1.5\times 10^{-3}.$  
This behavior follows naturally from the fact that in the modified 
divergence constraint (eq.~(\ref{eqn:divu_ev})),
$\tgamma (\x)$ accounts not only for the varying water composition but
also for all the latent heat released during
phase transitions (see Appendix \ref{app:mod_cons}); 
therefore, local variations in $\tgamma (\x)$
are expected to be more important.

\begin{figure}[!ht]
\centering
\includegraphics[width=0.49\textwidth]{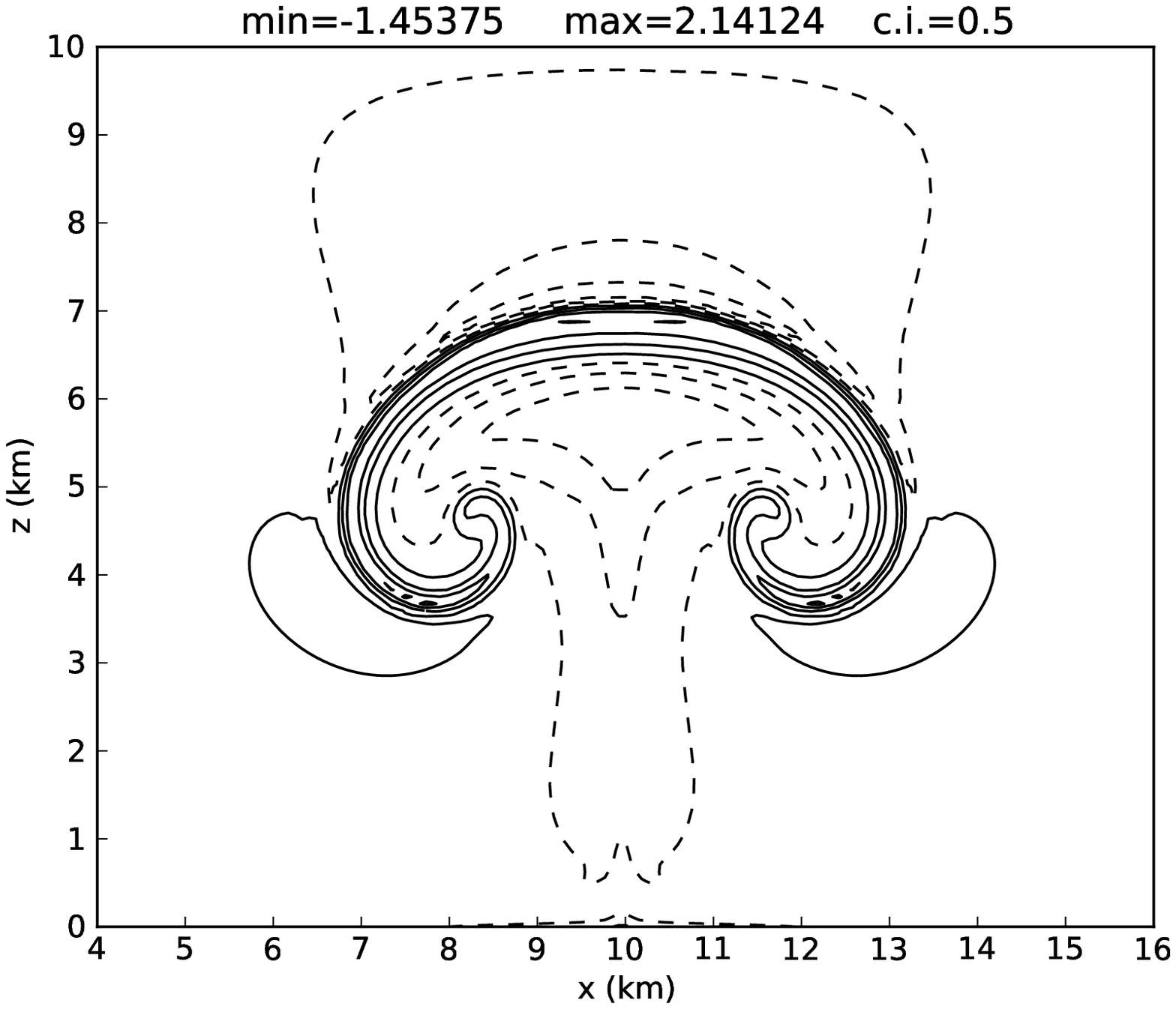}
\includegraphics[width=0.49\textwidth]{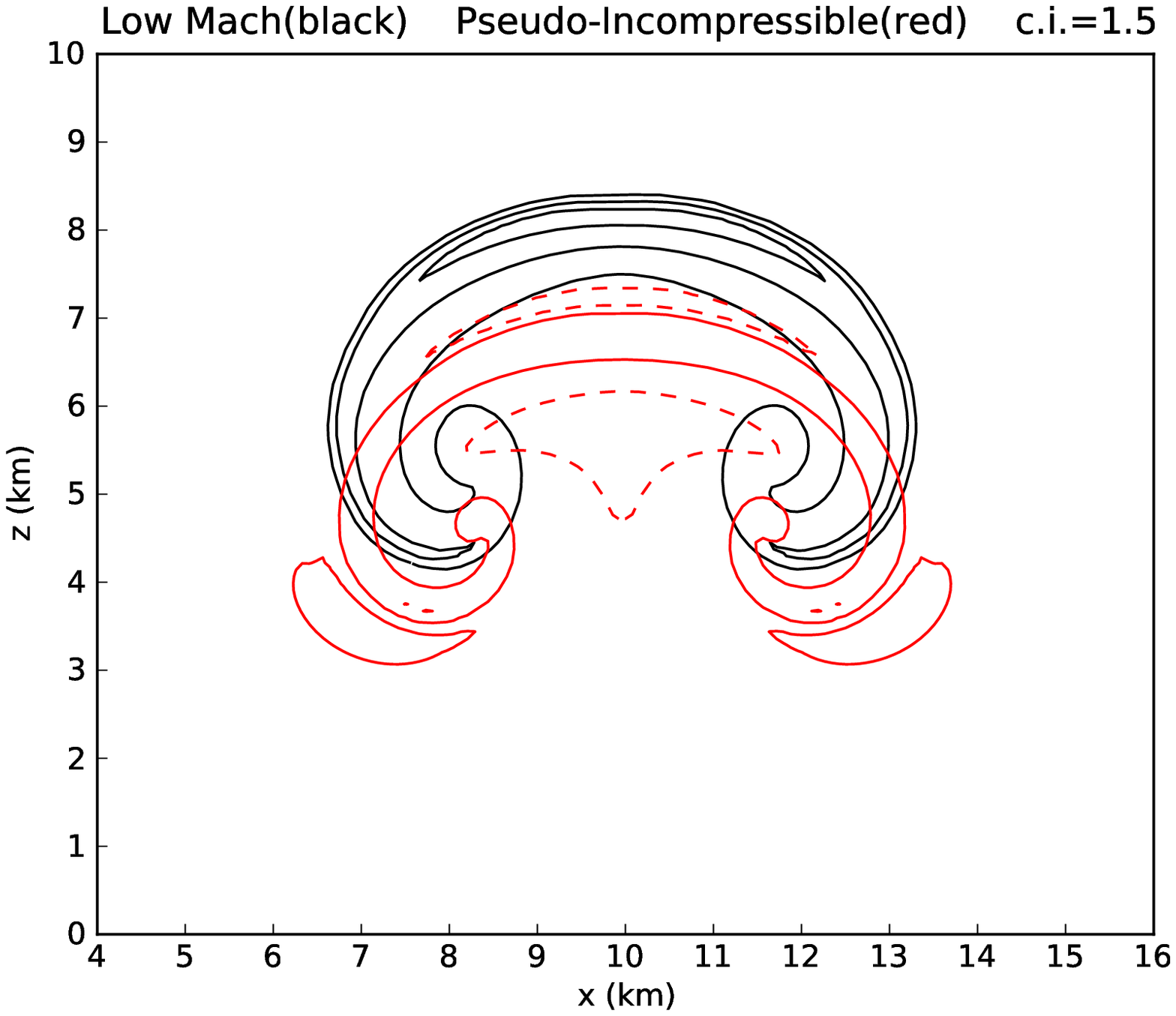}
\caption{Solution to the isentropic background problem neglecting
the specific heats of liquid water and vapor water.
Left: perturbational potential temperature shown with contours every $0.5\,$K;
negative contours are dashed.
Contrast to Fig.~6(a) in \cite{Bryan2002}.
Right: comparison with the low Mach number solution shown in 
Figure \ref{fig:compressible_comp} (bottom left);
perturbational potential temperature is shown with contours every $1.5\,$K.
}
\label{fig:pseudo_comp}
\end{figure}

The formalism adopted for the present low Mach model considers
straightforwardly the effects of the specific heats of liquid water 
and water vapor in the evaluation of the thermodynamic
properties of the moist fluid, and in particular in the definition of the
internal energy and specific enthalpy of moist air,
(\ref{eqn:ehat}) and (\ref{eqn:hhat}).
The latter is not possible within the moist pseudo-incompressible
model introduced in \cite{ONeill2013}, which relies on a 
$\theta - \pi$ 
(potential temperature -- Exner pressure) formalism defined
with the specific heat of dry air.
Neglecting the specific heats of liquid water 
and water vapor in our model amounts to consider the 
Equation set A investigated in \cite{Bryan2002} for the
same isentropic background configuration.
This is also the benchmark problem considered in 
\cite{ONeill2013} to validate their model.
Figure~\ref{fig:pseudo_comp} shows the low Mach number
solution when the specific heats of water are neglected
and contrasts it to the general case where all specific
heats are taken into account, like in Figure~\ref{fig:compressible_comp}. 
In \cite{Bryan2002},
the maximum and minimum values for 
$\theta'_e$ are 
$2.13647\,$K and $-1.39627\,$K, respectively.
As discussed in \cite{Bryan2002}, the impact of approximating
thermodynamic properties of moist air can lead to important 
variations for certain configurations, as illustrated in this example.

\subsection{Non--isentropic background state}\label{sec:non-isen}
We next consider a non--isentropic background state
given by the hydrostatically balanced profiles
in \cite{Clark1984} (eq.~2).  For the following computations
we define a computational domain $4\,$km high and wide, with periodic horizontal 
boundary conditions.  
The boundary conditions at the top and bottom boundaries are as described
for the isentropic case,
except for the thermodynamic variables which 
are extrapolated to determine the corresponding fluxes.
Again we use the configuration and parameters as given in \cite{CASTROmoist}.
All simulations with the low Mach formulation were performed on a uniform
grid of $256\times 256$.  As before, the reference compressible solutions 
were computed on a finer grid, here $1024\times 1024$.

The initial distributions of water vapor and liquid water 
in the atmosphere are set by the
relative humidity in the atmosphere, RH,
measured in percentage
and defined as RH$~= (p_v/\pvstar)\times 100$.
In particular if RH$_0 <100\,$\%,
then no liquid water should be initially present in the 
atmosphere in order to guarantee the thermodynamic 
equilibrium of the initial state.  Following \cite{CASTROmoist},
we consider in this study two cases:
first, a saturated medium with RH$_0 =100\,$\%
and $q_l>0$ everywhere in the domain as in the
isentropic background problem; 
and a second configuration with RH$_0 =20\,$\%,
and hence, no liquid water in the initial background state.

\begin{figure}[!ht]
\noindent
\includegraphics[width=0.32\textwidth]{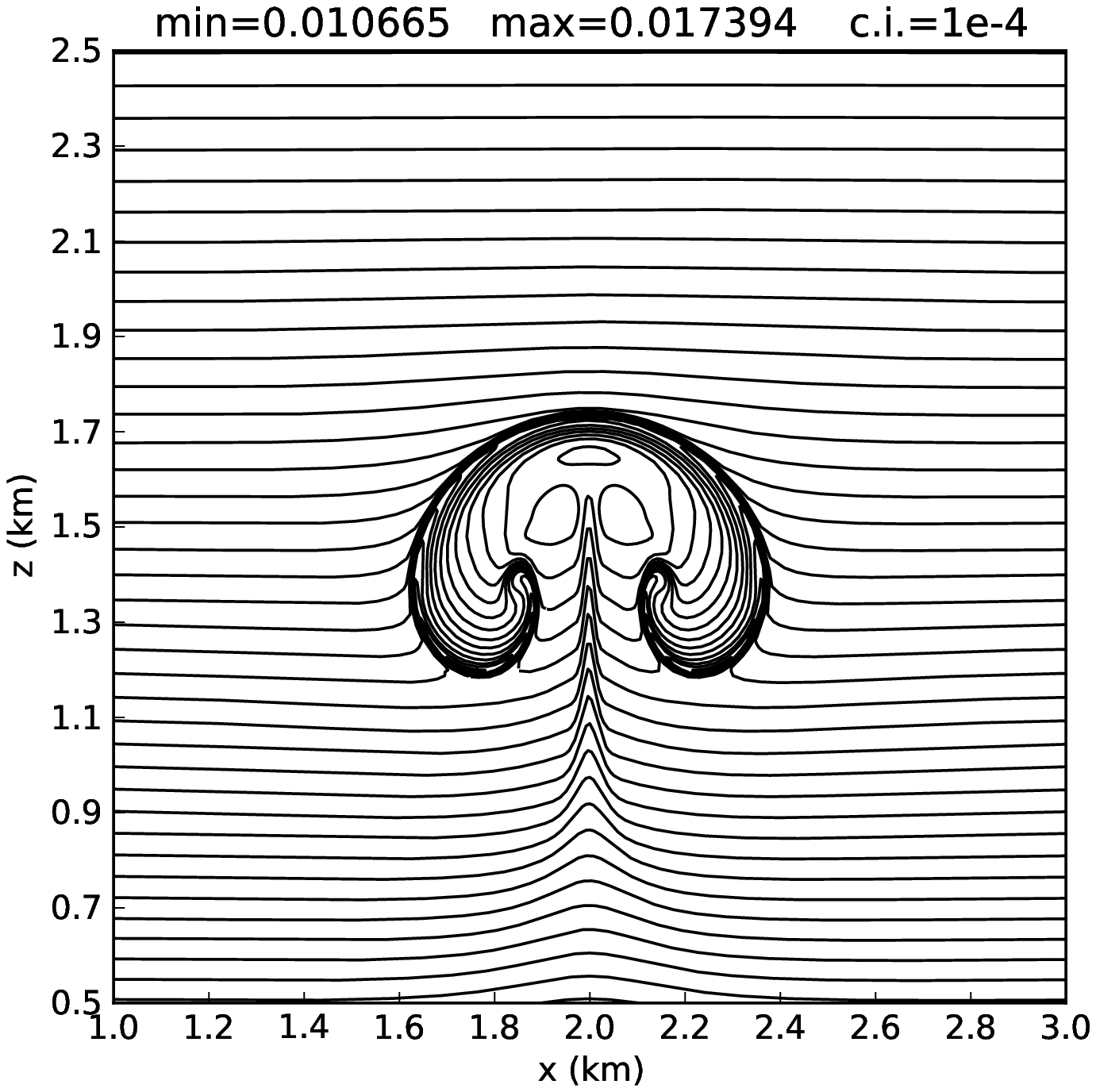}
\includegraphics[width=0.32\textwidth]{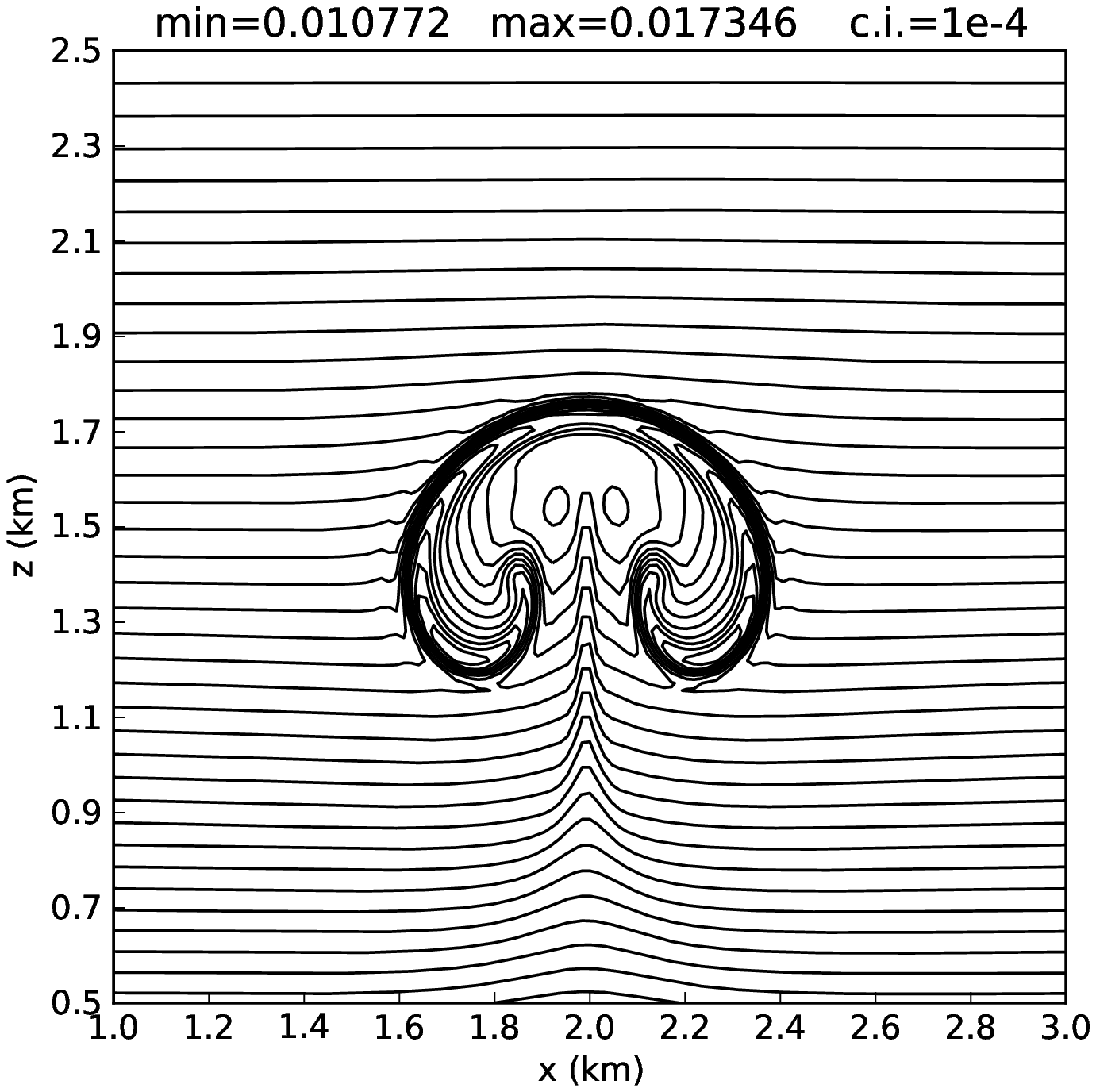}
\includegraphics[width=0.32\textwidth]{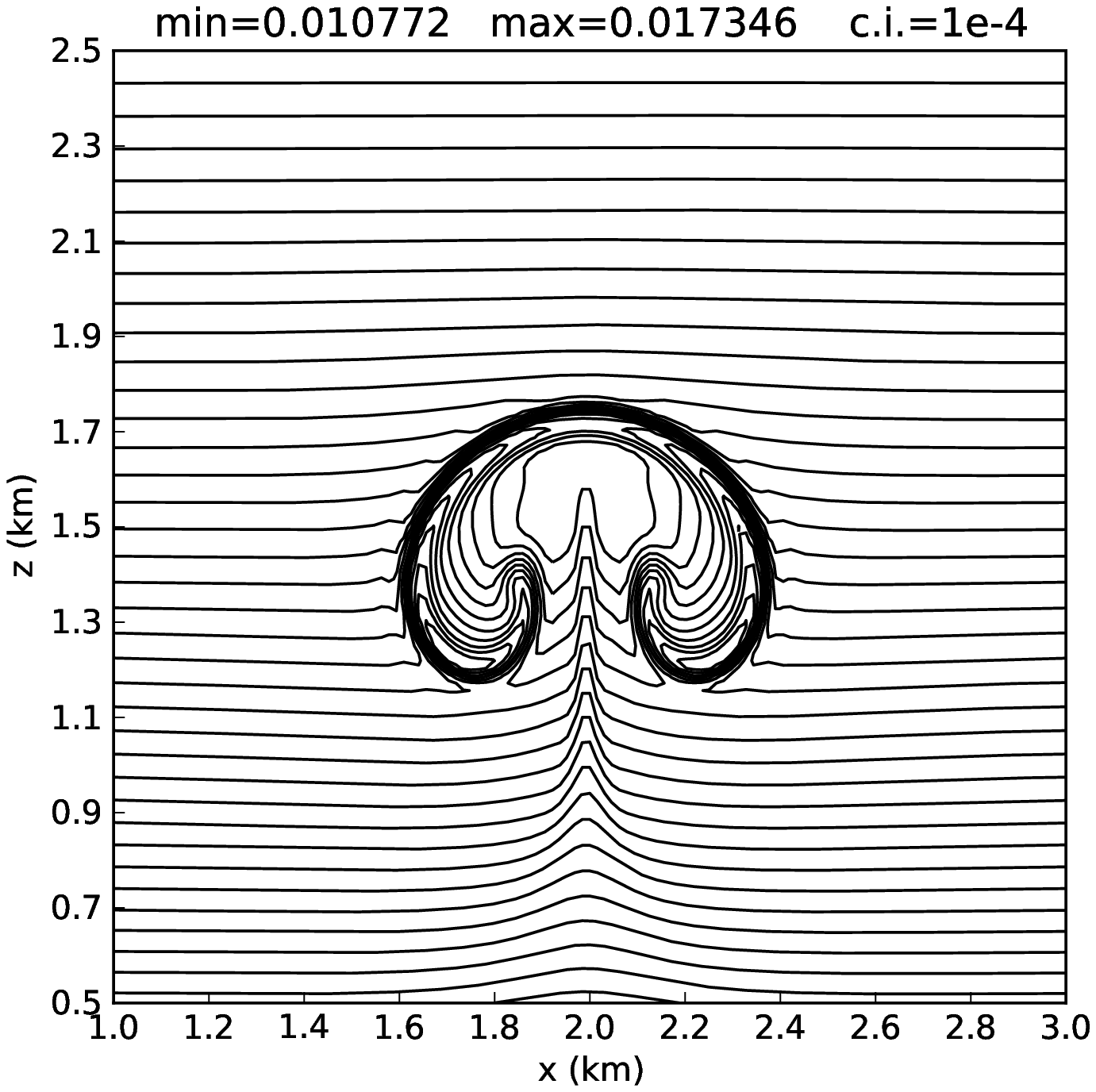}\\
\includegraphics[width=0.32\textwidth]{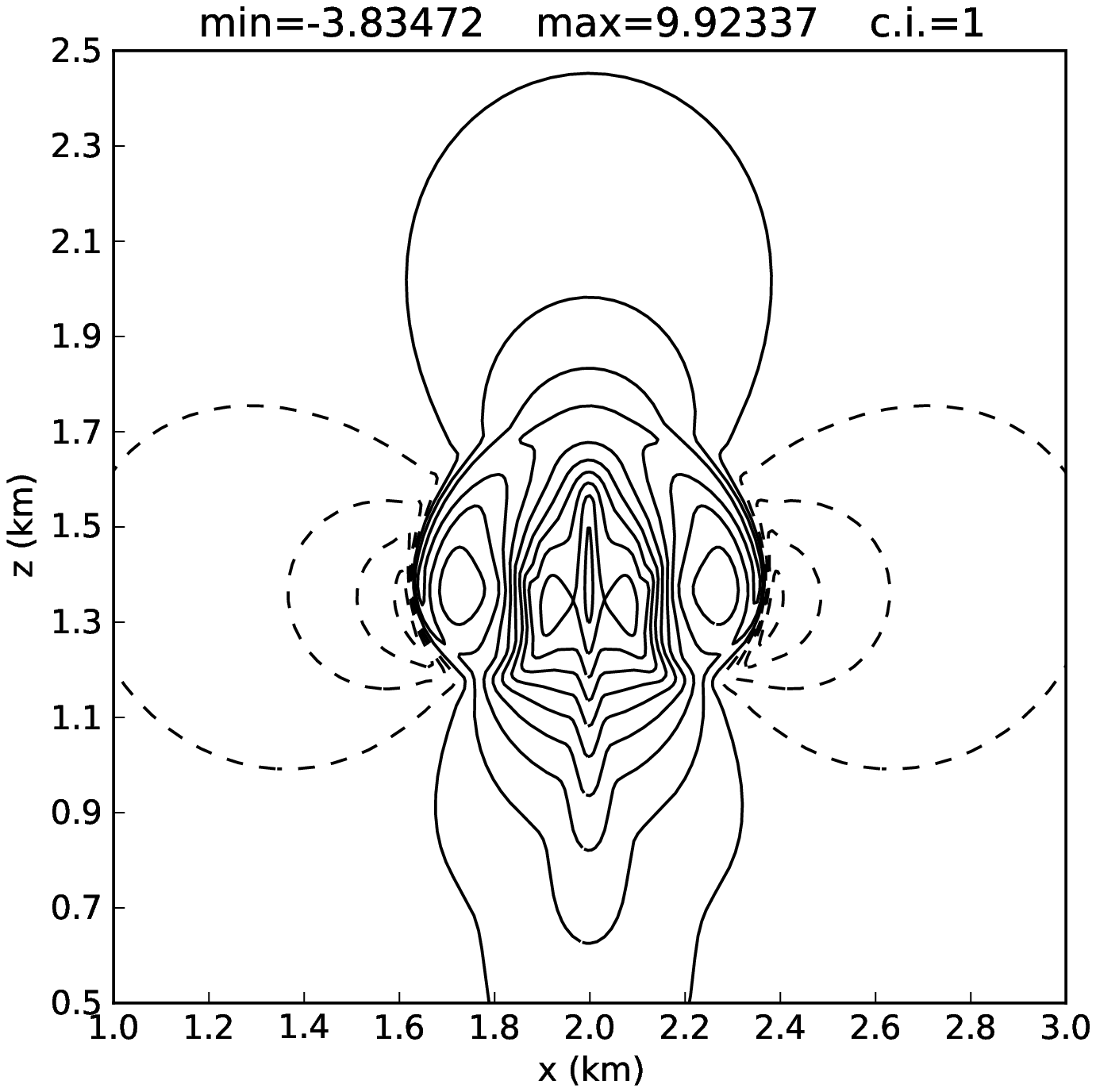}
\includegraphics[width=0.32\textwidth]{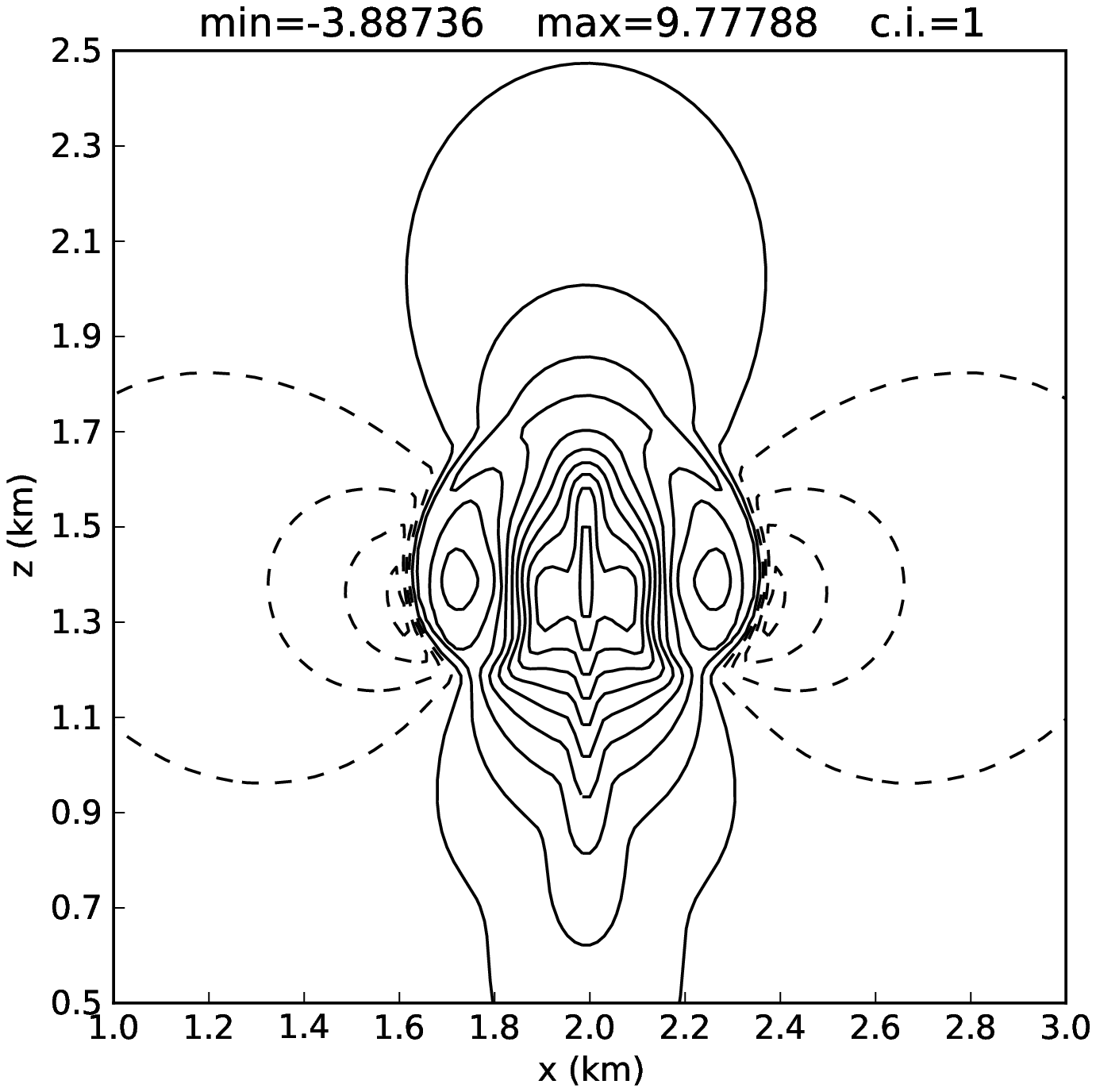}
\includegraphics[width=0.32\textwidth]{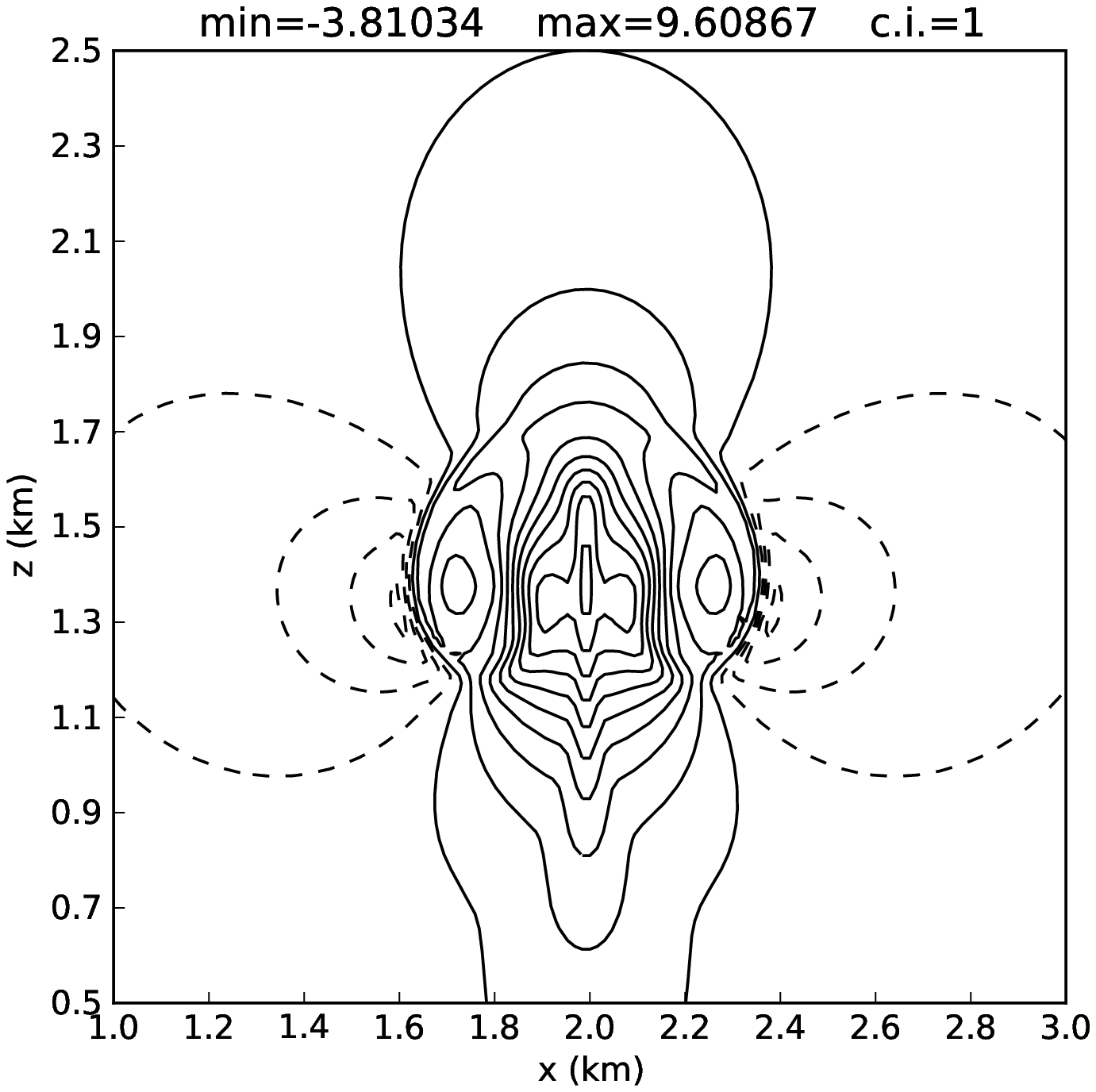}
\caption{Initially saturated, non--isentropic background state.
Liquid water mass fraction (top) and vertical velocity (bottom) are shown
for the reference compressible (left), and low Mach number solution at $300\,$s
using the first (center) and second (right) approach.
Contours are every $10^{-4}$ (top) and $1\,$m s$^{-1}$ (bottom);
negative contours are dashed.}
\label{fig:Grabowski_satmoist}
\end{figure}

Let us consider the first configuration 
with an initially saturated environment.
As described in \cite{CASTROmoist},
we initially introduce a warm perturbation of temperature.
Figure~\ref{fig:Grabowski_satmoist}
shows solutions obtained with the low Mach formulation
using the first and second approach for the divergence constraint,
as well as the compressible reference solution.
Solutions are very similar in all three cases
even though the low Mach approximations yield thermals
rising slightly faster.  In contrast to the previous
example, introducing the $\delta \Gamma_1$--correction  
does not visibly change the results.
For a $256\times 256$ grid and a CFL factor of $0.9$, 
the time steps for the low Mach approximations are about
$1.4\,$s compared to $0.04\,$s with the compressible formulation. 
The low Mach number simulation takes roughly a 
factor of $13$ less computational time than the compressible simulation.

\begin{figure}[!ht]
\centering
\includegraphics[width=0.49\textwidth]{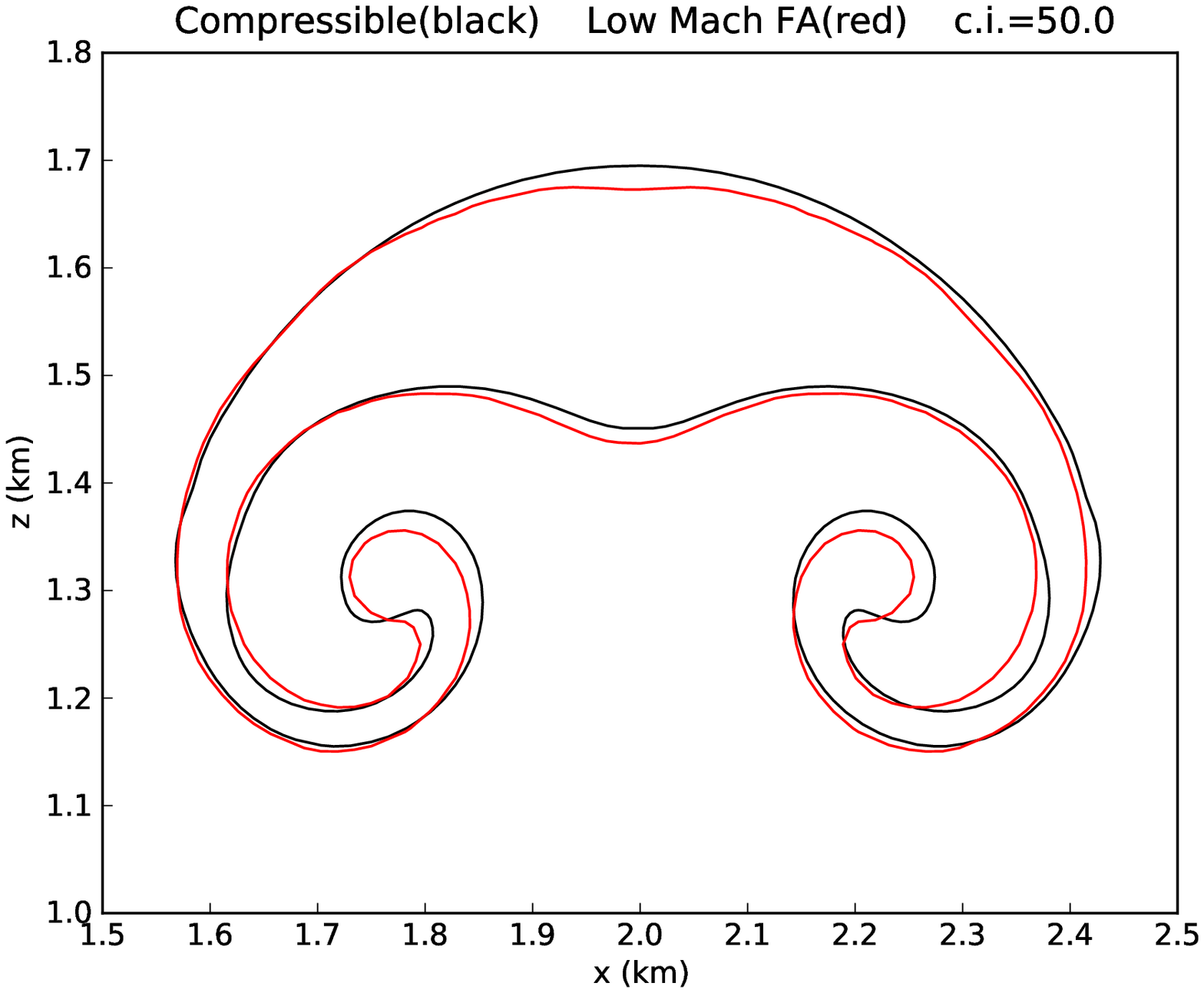}
\includegraphics[width=0.49\textwidth]{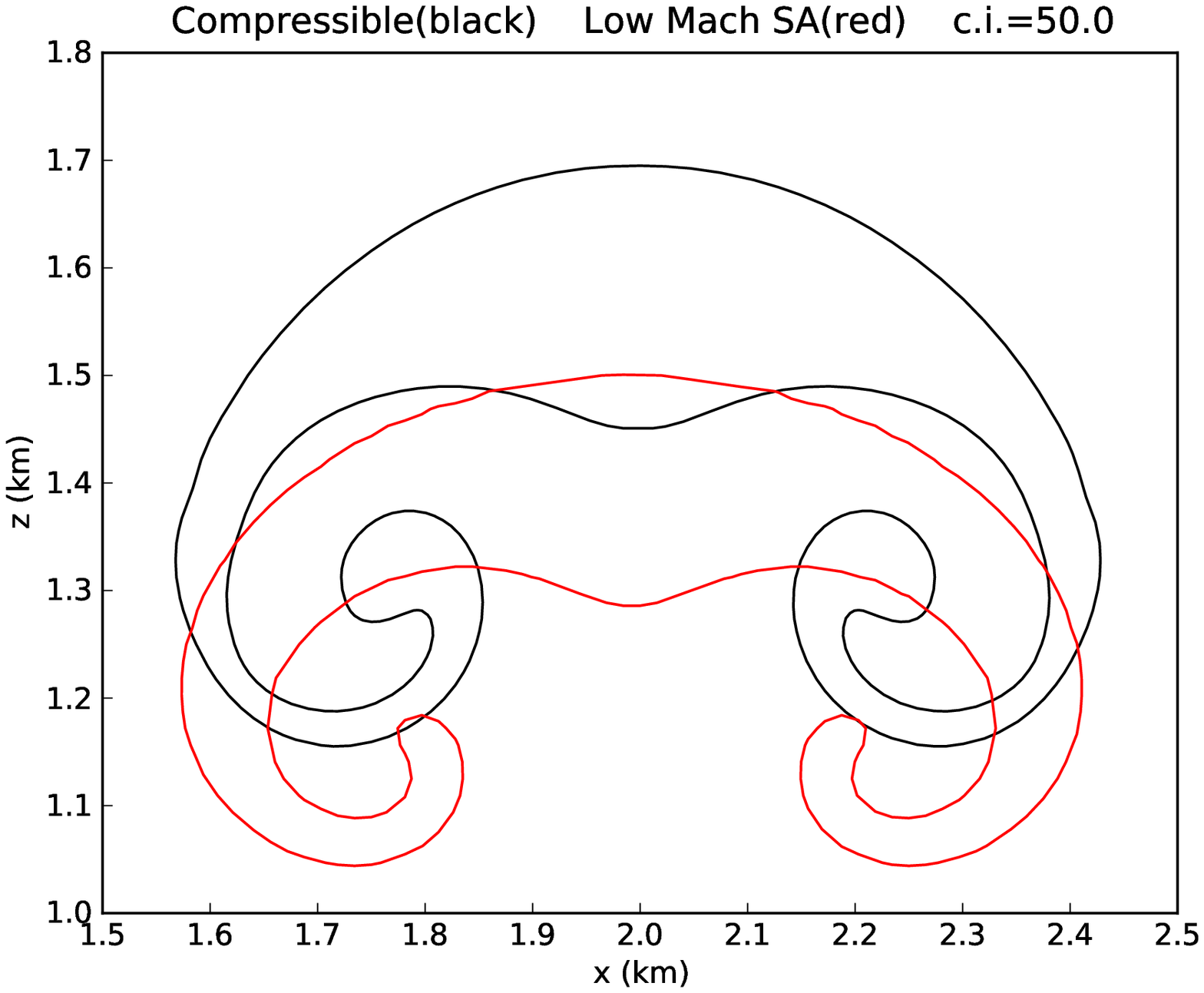}
\includegraphics[width=0.49\textwidth]{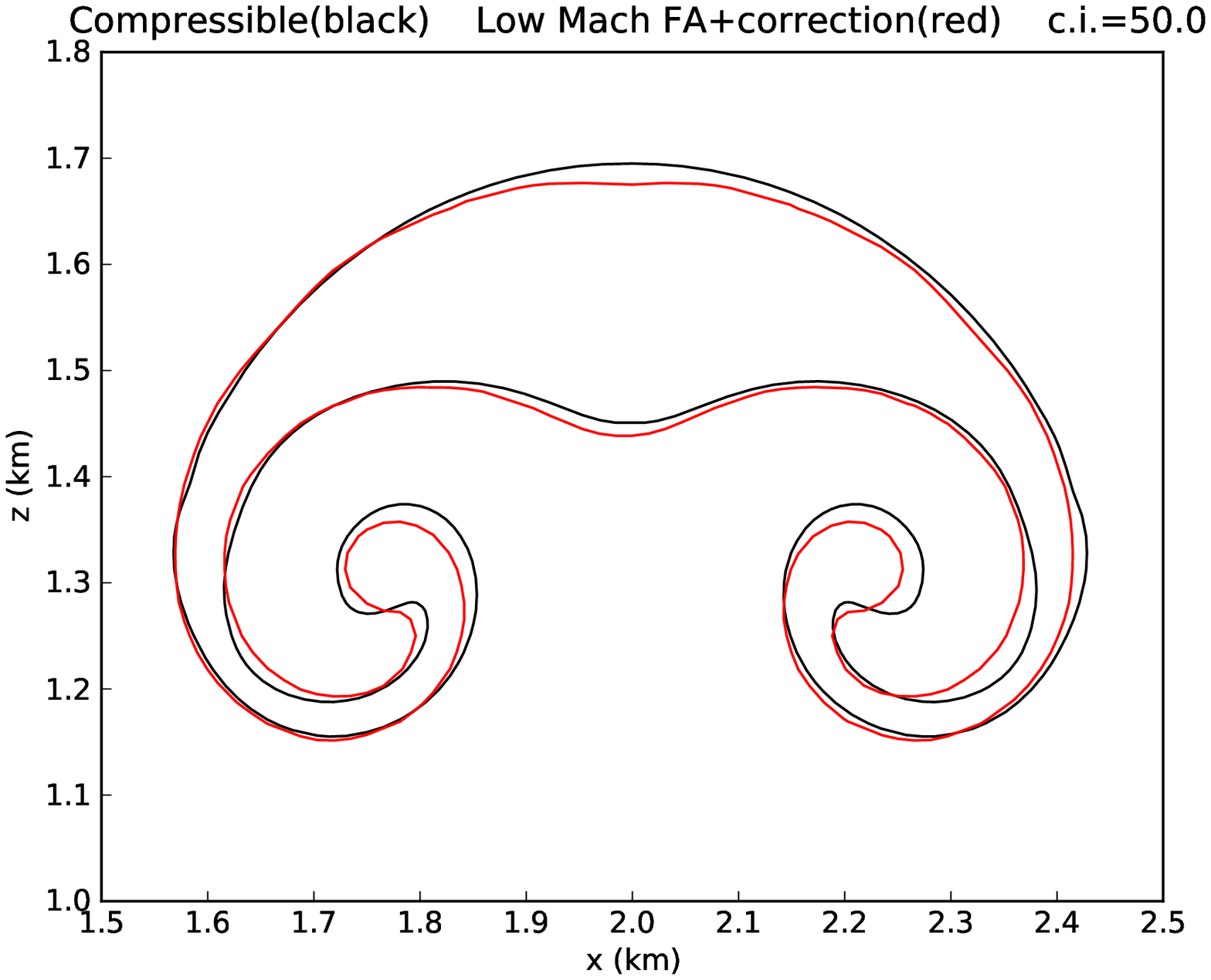}
\includegraphics[width=0.49\textwidth]{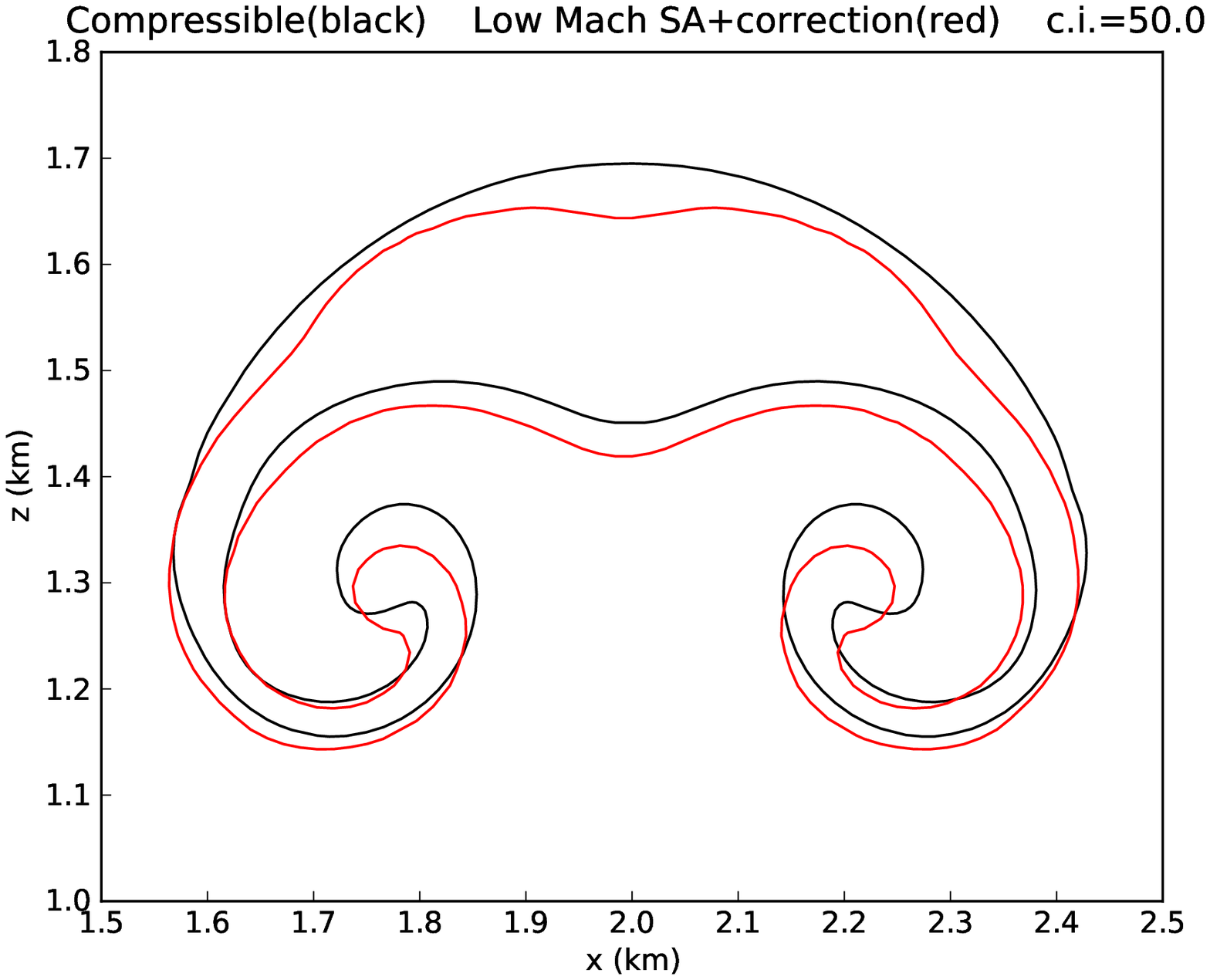}
\caption{Non--isentropic background state with a saturated perturbation.
Moist specific entropy is shown with contours every $50\,$J kg$^{-1}$ K$^{-1}$.
The low Mach number solution at $300\,$s (red)
using the first (left) and second (right) approach overlays the reference compressible solution (black).
On the bottom are simulations that use the $\delta \Gamma_1$--correction.
}
\label{fig:Grabowski_nnsatmoist}
\end{figure}

For the second configuration with RH$_0 =20\,$\%,
we consider the same previous temperature perturbation
and an additional circular perturbation 
in the relative humidity, which is set to $100\,$\%, 
with a transition layer, as detailed in \cite{CASTROmoist}.
Initially there is no liquid water in the domain.
Like Figure~\ref{fig:compressible_comp}, Figure~\ref{fig:Grabowski_nnsatmoist} 
compares the two low Mach number solutions to the reference compressible solution;
the top row shows the results using laterally averaged $\gamma_m$ and $\tgamma,$ while
the bottom row shows the results using the $\delta \Gamma_1$--correction.
Here the deviation of $\gamma_m(\x)$ from $\gammambar (z)$ in the first approach 
ranges from $-4.2\times 10^{-3}$ to $6.9\times 10^{-4},$
which is considerably smaller than 
the deviation of $\tgamma (\x)$ from $\tgammambar (z)$ in the second
approach, which ranges from $-0.17$ to $3.3\times 10^{-2}$.
\begin{figure}[!ht]
\centering
\includegraphics[width=0.49\textwidth]{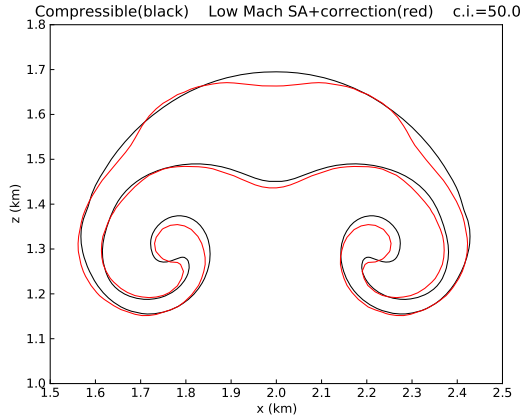}
\caption{Non--isentropic background state with a saturated perturbation.
Moist specific entropy is shown with contours every $50\,$J kg$^{-1}$ K$^{-1}$.
The low Mach solution at $300\,$s (red)
using the $\delta \Gamma_1$--correction given by (\ref{eq:gammafull_2ndorder})
overlays the reference compressible solution (black).}
\label{fig:Grabowski_nnsatmoist_2ndorder}
\end{figure}

Better agreement can be seen in Figure~\ref{fig:Grabowski_nnsatmoist_2ndorder},
where a term of order $\delta \Gamma_1^2$
is also considered in (\ref{eq:gammafull}) for the 
$\delta \Gamma_1$--correction:
\begin{equation}\label{eq:gammafull_2ndorder}
\Dx \cdotb \vel = 
- \frac{1}{\gammabar \po} \frac{D\po}{Dt}
+ S  +
\frac{\delta \Gamma_1}{\gammabar^2 \po} 
\frac{D\po}{Dt}
- \frac{\delta \Gamma_1^2}{\gammabar^3 \po} 
\frac{D\po}{Dt}. 
\end{equation}
The time steps used in the low Mach computations are 
about $2.1\,$s, compared to $0.04\,$s with the 
compressible formulation. 
The low Mach number simulation takes roughly a 
factor of $15$ less computational time than the compressible simulation.

\subsection{Three--dimensional simulation}
Finally, we consider two interacting thermals
rising in a three--dimensional, non--isentropically
stratified background in a domain $10\,$km on a side
and $15\,$km high.
The background state is defined as in 
\S\ref{sec:non-isen},
with a relative humidity of RH$_0 =20\,$\% 
and no liquid water in the initial configuration.
The following temperature perturbation is then introduced:
\begin{equation*}\label{eqn:temp_pert3D}
T' = 6 \cos ^2 \left( \frac{\pi L_1}{2} \right)
   + 6 \cos ^2 \left( \frac{\pi L_2}{2} \right),
\end{equation*}
where 
$L_1 = \min (1, r_1), \;\;  L_2 = \min (1, r_2),$
and 
\begin{equation*}\label{eq:pert_L3D_1}
r_1 = \frac{1}{4} \sqrt{\left({x-x_1} \right)^2
+\left({y-y_1} \right)^2 + \left({z-z_1} \right)^2 } \; ,
\end{equation*}
\begin{equation*}\label{eq:pert_L3D_2}
r_2 = \frac{1}{3}\sqrt{\left({x-x_2} \right)^2
+\left({y-y_2} \right)^2 + \left({z-z_2} \right)^2 }, 
\end{equation*}
with 
$x_1 = y_1 = 5\,$km, $z_1 = 7.5\,$km
and
$x_2 = y_2 = 7\,$km,
$z_2 = 7.5\,$km.
Within the regions where the temperature is perturbed, we also
perturb the relative humidity by setting 
RH equal to $50\,$\% for 
$r_1 < 3\,$km and $r_2 < 2\,$km, 
for each initial thermal,
with corresponding transition layers:
\begin{equation*}\label{eqn:rh_pert3D_1}
 {\rm RH} = {\rm RH}_0 + (50 - {\rm RH}_0)
\cos ^2 \left( \frac{\pi }{2}
\left[r_1-3\right] \right),
\
3 \leq r_1 \leq 4,
\end{equation*}
\begin{equation*}\label{eqn:rh_pert3D_2}
 {\rm RH} = {\rm RH}_0 + (50 - {\rm RH}_0)
\cos ^2 \left( \frac{\pi }{2}
\left[r_2-2\right] \right),
\
2 \leq r_2 \leq 3.
\end{equation*}
\begin{figure}[!ht]
\centering
\includegraphics[width=0.49\textwidth]{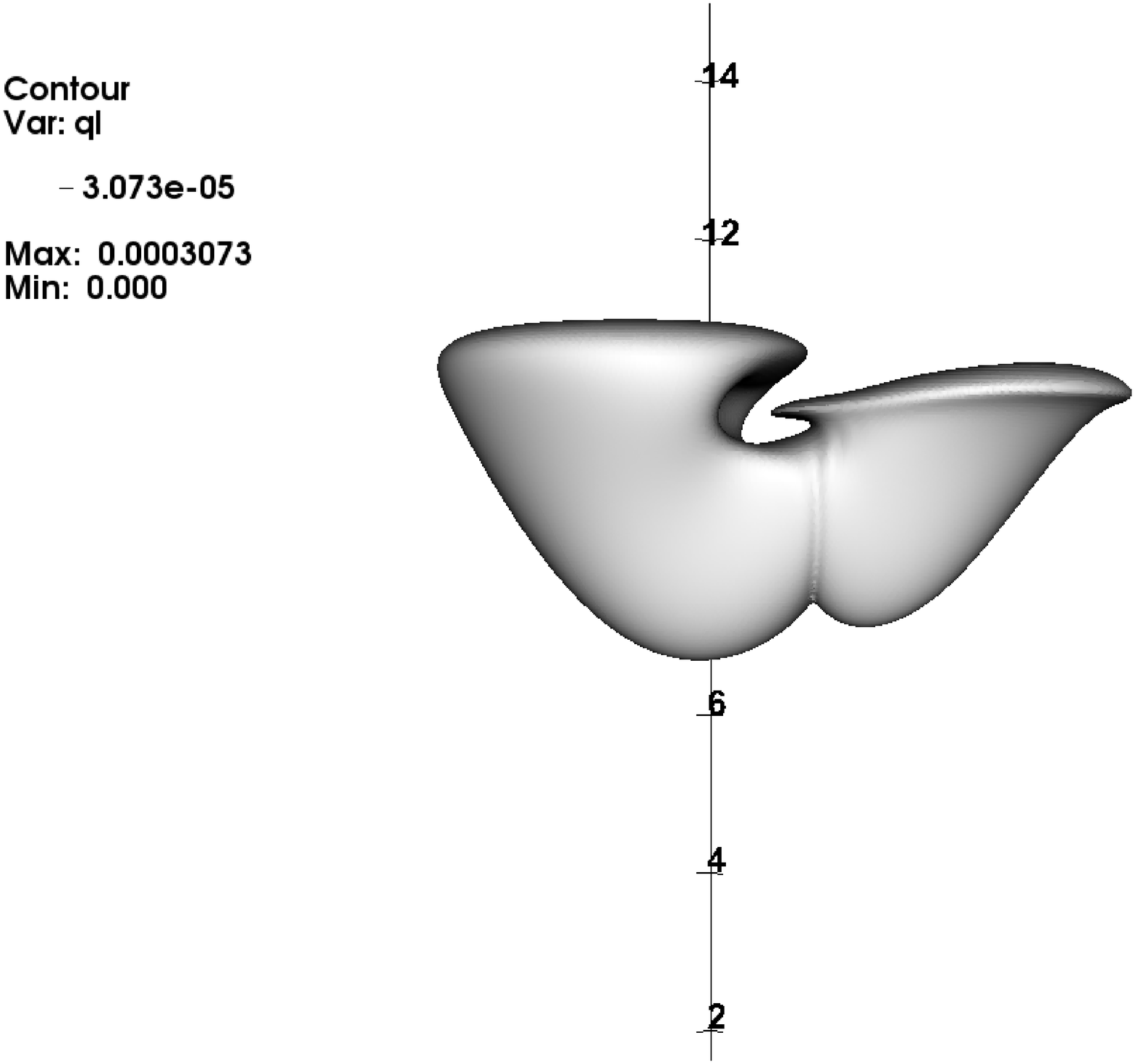}
\includegraphics[width=0.49\textwidth]{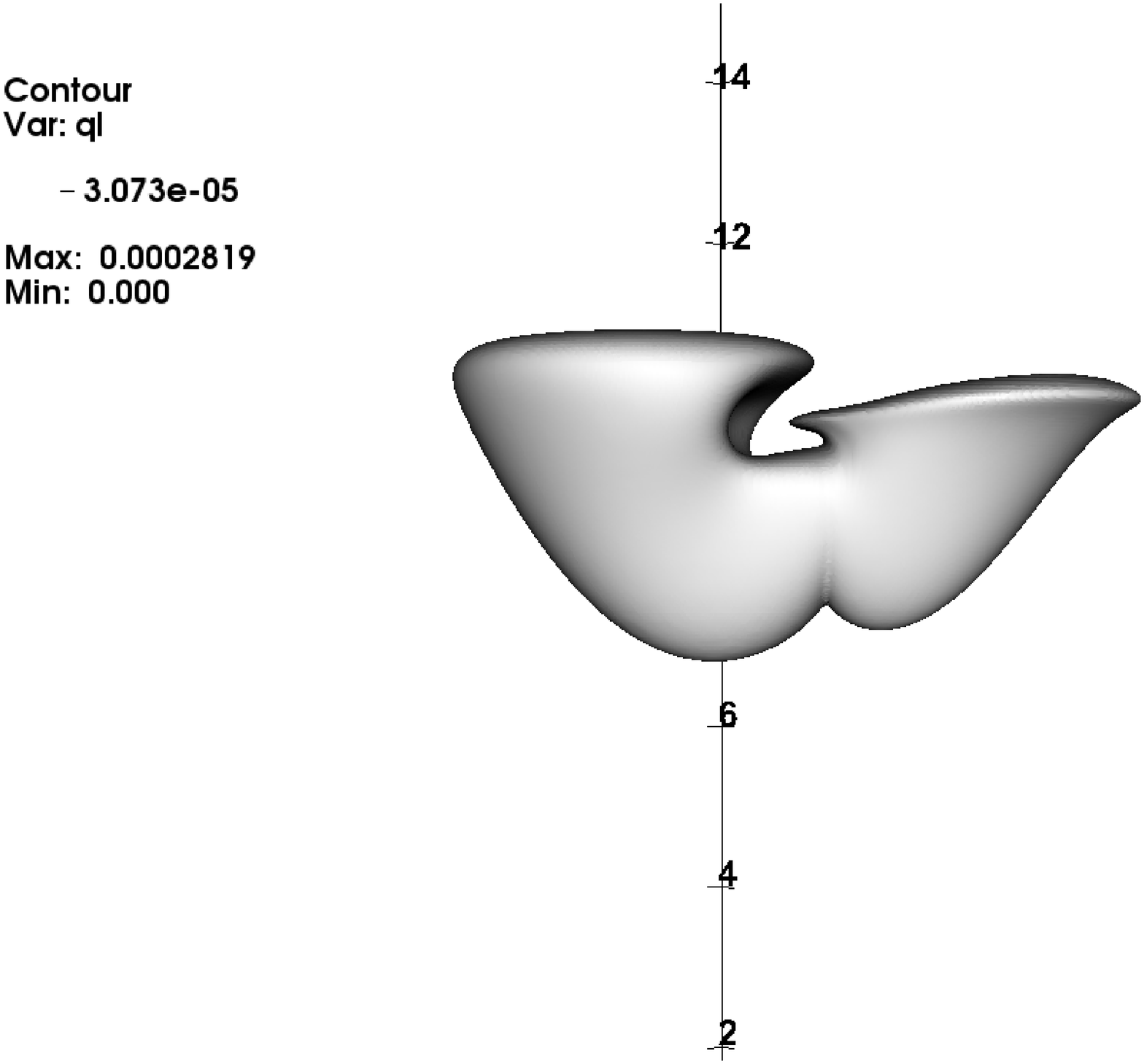}
\includegraphics[width=0.49\textwidth]{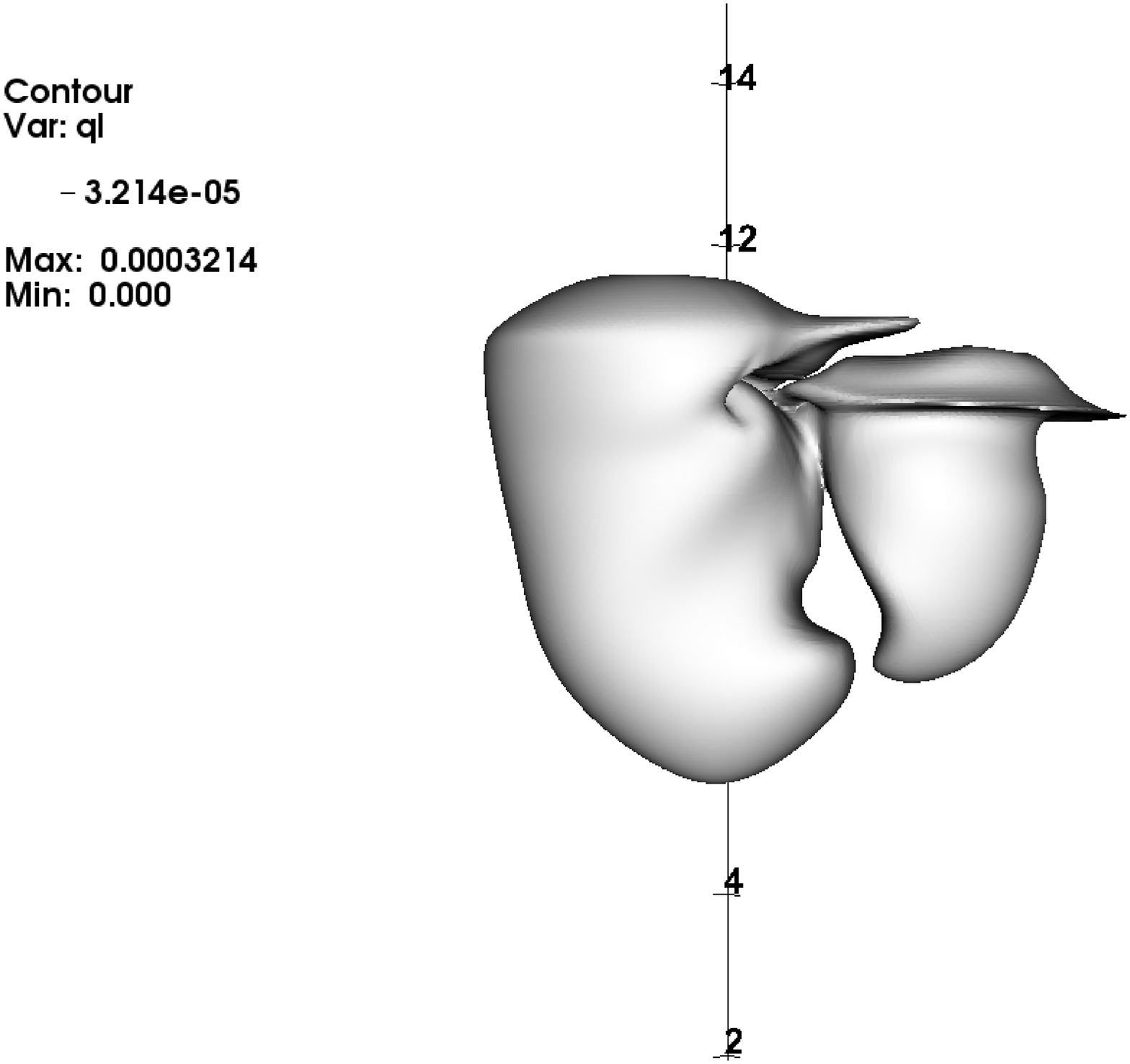}
\includegraphics[width=0.49\textwidth]{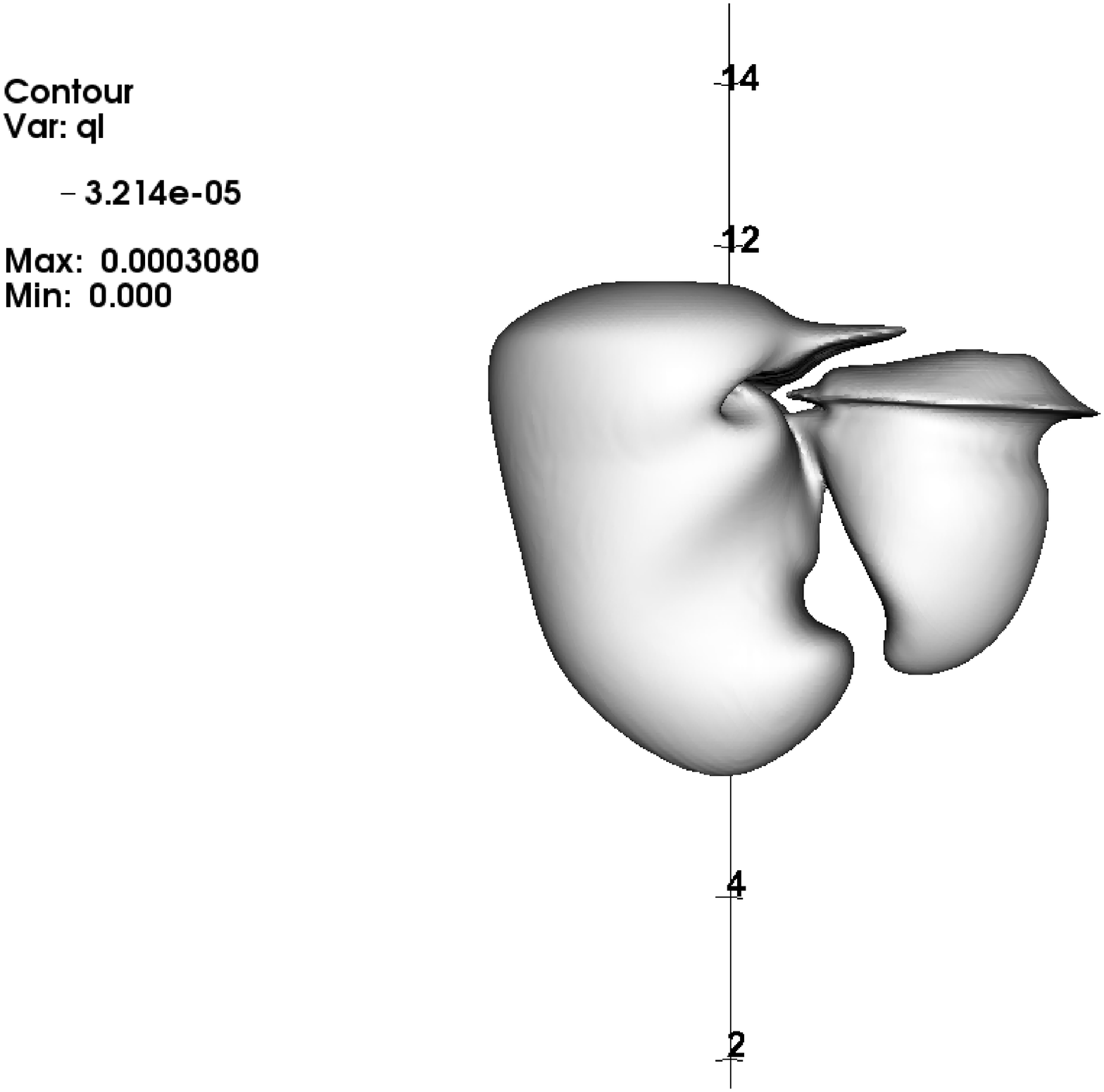}
\caption{Two interacting thermals in a three--dimensional 
non--isentropic background state.
Isosurfaces of liquid water for the reference compressible solutions
(left) and the low Mach number ones (right)
at times $500\,$s 
($q_l=3.073\times 10^{-5}$)
(top) and $1000\,$s (bottom)
($q_l=3.214\times 10^{-5}$).
The low Mach number solver uses the modified divergence constraint
(\ref{eqn:divu_ev}) (second approach)
with the $\delta \Gamma_1$--correction given by (\ref{eq:gammafull_2ndorder}).}
\label{fig:Grabowski_3D}
\end{figure}
\begin{figure}[!ht]
\centering
\includegraphics[width=0.49\textwidth]{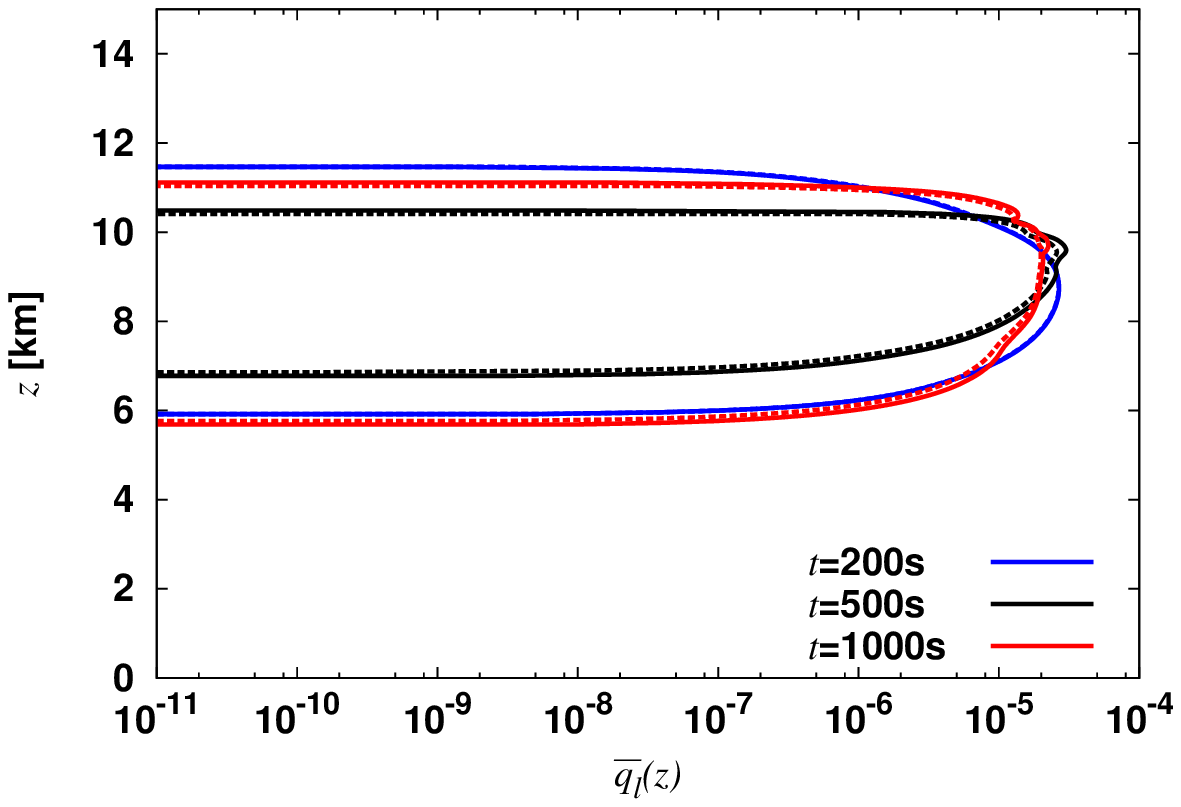}
\includegraphics[width=0.49\textwidth]{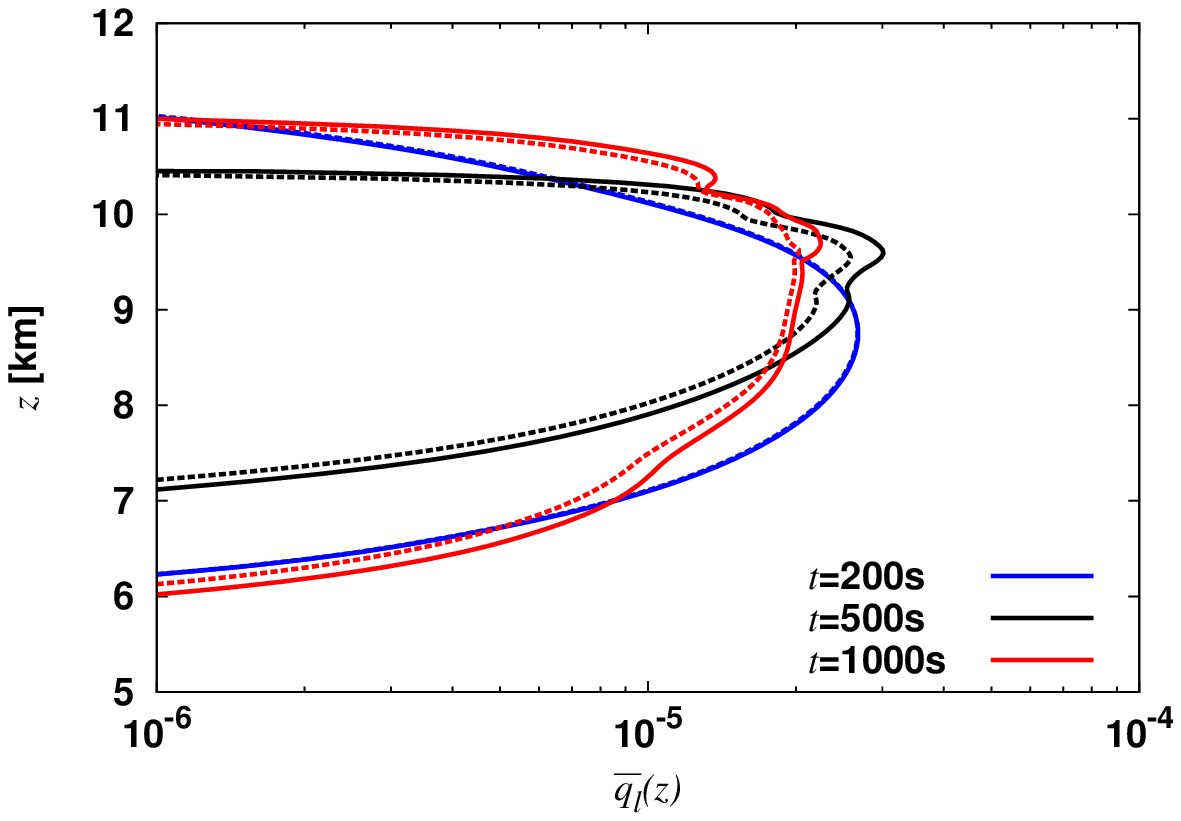}
\caption{Horizontal budgets of liquid water, $\overline{q_l}(z)$,
computed with (\ref{eq:lateral_average}) at $t = 200\,$s, $500\,$s and $1000\,$s.
Reference compressible and low Mach number
solutions are represented with solid and dashed lines, respectively.
More details can be appreciated in the zoomed region (right).}
\label{fig:budget_ql}
\end{figure}

We consider the low Mach number formalism using the second
approach (modified divergence constraint (\ref{eqn:divu_ev}))
to numerically implement phase transitions,
with the $\delta \Gamma_1$--correction given by (\ref{eq:gammafull_2ndorder}).
As before, the low Mach number solutions are contrasted 
to reference compressible solutions.  For a uniform grid of 
size $256 \times 256 \times 384$, the time step in the low Mach number simulation 
is approximately $3\,$s, compared to $0.1\,$s in the compressible simulation.
For this particular problem, the total run time of the 
low Mach number simulation is roughly a factor of $5$ less
than that of the compressible simulation.
Figure \ref{fig:Grabowski_3D} illustrates the formation
of liquid water as computed with both formulations.
The relative difference between the maximum values of $q_l$ in 
the compressible and the low Mach number solutions is roughly $8\,$\%
at $t = 500\,$s and $4\,$\% at $t=1000\,$s, the two times shown in 
Figure \ref{fig:Grabowski_3D}.
Figure \ref{fig:budget_ql} shows the horizontal budgets of liquid water,
$\overline{q_l}(z)$, computed using formula (\ref{eq:lateral_average}),
at $t = 200$s and at the two simulation times  
shown in Figure \ref{fig:Grabowski_3D}.
For $t = 200\,$s practically the same solution is recovered
with both formulations, with a
relative difference of $0.2\,$\% between 
the maximum values of $q_l$.
Recalling that $q_l$ is diagnostically recovered in both
the compressible and the low Mach number approach,
we can track the formation of liquid water by computing it
after each time step for comparison purposes.
Local values of $q_l$ larger than $10^{-10}$ appear
after $111\,$s and $112.5\,$s for the compressible
and low Mach number simulations, respectively.
The Mach number $M$ associated with this particular problem
remains lower than $0.05$ during the entire numerical simulation.

\section{Summary}
We have presented a new low Mach number model for moist 
atmospheric flows with a general equation of state,
based on the low Mach number model for stratified reacting flows
introduced in \cite{ABNZ:III}.  
In our model we consider only reversible processes, 
namely water phase changes as in \cite{ONeill2013}, 
using an exact Clausius--Clapeyron formula for moist thermodynamics 
and considering the effects of the specific heats of water
and the temperature dependency of the latent heat.
A set of invariant variables
was used as prognostic variables in the equations of motion, 
including in particular 
the total water content and a specific enthalpy of moist air
that accounts for the contribution of both sensible and latent heats.
The evolution equations 
can thus be solved without needing 
to estimate or neglect source terms related to phase change during
the time integration.
The mass fractions of water vapor and liquid water 
are diagnostically recovered as required during a time step
by imposing the saturation requirements of an atmosphere at thermodynamic
equilibrium.  The latter is an important property since 
updating 
the solution while
ignoring the varying water composition,
may  negatively
impact the accuracy of the moist flow dynamics,
as investigated in \cite{CASTROmoist}.

We then considered a moist thermodynamic model that treats
dry air and water vapor as ideal gases to define the
equation of state for moist air.
In order to account for the latent heat release
in the low Mach divergence constraint for the velocity field,
the evaporation rate is estimated from the time variation
of saturated water vapor within a parcel.
The amount of saturated water vapor within a parcel
is determined by the 
Clausius--Clapeyron formula as a function of the 
local thermodynamical state; 
the evolution of the state depends on the local advected motions.
An analytical expression for the evaporation rate was thus
derived that depends on local parameters given by 
the temperature and pressure, the water composition,
and the velocity field.  Two approaches were then considered.
In the first, the rate of phase change can be computed to 
evaluate the latent heat release; in the second, 
a modified divergence constraint can be analytically 
deduced by introducing the derived expression for the evaporation rate 
in the original divergence constraint.
Both approaches are analytically equivalent and together
with the low Mach number equation set allow us to 
characterize moist atmospheric flows.

The MAESTRO code\footnote{Available at 
http://bender.astro.sunysb.edu/Maestro/download/} 
\cite{multilevel}, originally designed to simulate
stratified reacting flows arising in astrophysical settings, was 
adapted to model moist atmospheric flows.
A series of test problems was investigated 
with both isentropic and non--isentropic background states,
as well as saturated and partially saturated regions in the 
atmosphere.  Results were contrasted to 
reference solutions obtained with a fully compressible 
formulation.  Very good agreement with the reference
moist dynamics was shown using both the first and second approach
(with the $\delta \Gamma_1$--correction),  
thus demonstrating that low Mach number models can serve as a reasonably 
accurate and computationally efficient alternative to compressible
codes for small-scale moist atmospheric applications.

\section*{Acknowledgments} 
This material is based upon work supported by the 
U.S. Department of Energy, Office of Science, 
Office of Advanced Scientific Computing Research, 
Applied Mathematics program under contract number DE-AC02005CH11231.

\appendix

\section{Derivation of the low Mach divergence constraint}\label{app:LM_constraint}
We rewrite the conservation of mass (eq.~(\ref{eqn:cont})) as
an expression for the divergence of velocity,
\begin{equation}\label{eqn:divu}
\Dx \cdotb \vel = -\frac{1}{\rho} \frac{D \rho}{D t}.
\end{equation}
Differentiating the equation of state, written as
$p=p(\rho,T,q_a,q_v,q_l)$, along particle paths, we obtain
\begin{align*}\label{eqn:dpdt}
\frac{D \rho}{D t}
& = 
\frac{1}{p_{\rho}}
\left(
\frac{D p}{D t}
-p_{T} \frac{D T}{D t} 
- \sum_{i \in (a,v,l)}
p_{q_i}
\frac{D q_i}{D t}
\right)
\nonumber \\
&  
= 
\frac{1}{p_{\rho}}
\left(
\frac{D p}{D t}
-p_{T} \frac{D T}{D t} 
- \left(p_{q_v} - p_{q_l} \right) \frac{e_v}{\rho}
\right),
\end{align*}
with 
$p_{\rho} = \partial p/\partial \rho \vert_{T,q_i}$,
$p_{T} = \partial p/\partial T \vert_{\rho,q_i}$,
and 
$p_{q_i} =  \partial p/\partial q_i\vert_{\rho,T,(q_j,j\neq i)}$.
An expression for $DT/Dt$ can be obtained by differentiating
the definition of moist enthalpy (eq.~(\ref{eqn:hhat})),
and comparing terms with the enthalpy equation (\ref{eqn:enthalpy}):
\begin{align*}
\rho\frac{D \hhat}{D t} & =
\rho \left(
\left. 
\frac{\partial \hhat}{\partial T} \right\vert_{p,q_i}
\frac{D T}{D t} + 
\left.
\frac{\partial \hhat}{\partial p} \right\vert_{T,q_i}
\frac{D p}{D t} + 
\sum_{i \in (a,v,l)}
\left. 
\frac{\partial \hhat}{\partial q_i} \right\vert_{T,p,(q_j,j\neq i)}
\frac{D q_i}{D t}
\right)\nonumber\\
& =
\rho \left(
\cpm \frac{D T}{D t} + 
\hhat_p
\frac{D p}{D t} + 
\left(\hhat_{q_v} - \hhat_{q_l} \right) \frac{e_v}{\rho}
\right)\nonumber\\
& =
\frac{D p}{D t} + \rho \Hhat,
\end{align*}
or, gathering terms,
\begin{equation}\label{eqn:DTDthhat}
 \frac{D T}{D t} =
\frac{1}{\rho \cpm} \left[
\left(1- \rho\hhat_p\right) \frac{D p}{D t}
- \left(\hhat_{q_v} - \hhat_{q_l} \right) e_v
+ \rho \Hhat \right],
\end{equation}
where $\cpm = \partial \hhat/\partial T \vert_{p,q_i}$
is the specific heat of moist air at constant pressure,
$\hhat_p = \partial \hhat/\partial p \vert_{T,q_i}$,
and $\hhat_{q_i} = \partial \hhat / \partial q_i \vert_{T,p,(q_j,j\neq i)}$.
Coming back to equation (\ref{eqn:divu})
and replacing $p$ by $\po(z,t)$, 
we can write the divergence constraint on the velocity field as
(\ref{eqn:divu_alpha}) with 
$\alpha$ and $S$ given by (\ref{eqn:S}).

\section{Derivation of the evaporation rate}\label{app:evp_rate}
Considering 
that $e_v=\rho D \qvstar/Dt$ from
(\ref{eqn:ev_qvstar}) 
and differentiating $\qvstar = \qvstar(\rho,T)$ (eq.~(\ref{eqn:qvstar})) 
along particle paths,
we obtain 
\begin{equation*}\label{eqn:ev_DD_t}
e_v = \rho
\left[ 
\left. \frac{\partial \qvstar}{\partial \rho} \right\vert_{T}
\frac{D \rho}{D t} +
\left. \frac{\partial \qvstar}{\partial T} \right\vert_{\rho}
\frac{D T}{D t}
\right],
\end{equation*}
with
\begin{equation*}\label{eqn:dqvstar}
\left. \frac{\partial \qvstar}{\partial \rho} \right\vert_{T} =
- \frac{\qvstar}{\rho}, \qquad
\left. \frac{\partial \qvstar}{\partial T} \right\vert_{\rho} = 
\qvstar 
\left( \frac{\aav -1}{T} + \frac{\bbv}{ {T}^2 } \right) =  
\qvstar \phi (T),
\end{equation*}
according to (\ref{eqn:qvstar}) and (\ref{eqn:pvstar}).
With the equation of state (\ref{eqn:pres}),
equation (\ref{eqn:DTDthhat}) becomes
\begin{equation}\label{eqn:DTDthhatpres}
 \frac{D T}{D t} =
\frac{1}{\rho \cpm} \left[
\frac{D p}{D t}
- L_e e_v
+ \rho \Hhat \right],
\end{equation}
and hence,
\begin{equation}\label{eqn:ev_hhat_rho1}
e_v = 
\frac{\qvstar 
\left[
\rho \cpm \left(\Dx \cdotb \vel\right) + 
\phi (T) \ds
\left( \frac{D p}{D t} + \rho \Hhat \right)
\right] }
{\cpm + \qvstar \phi (T) L_e};
\end{equation}
that is,
\begin{equation}\label{eqn:ABC1}
\Aev  =
\frac{\qvstar 
\rho \cpm }
{\cpm + \qvstar \phi (T) L_e},
\qquad
\Bev =
\frac{\qvstar 
\phi (T) }
{\cpm + \qvstar \phi (T) L_e},
\qquad
\Cev =
\frac{\qvstar 
\phi (T) }
{\cpm + \qvstar \phi (T) L_e},
\end{equation}
into (\ref{eqn:ev_hhat_rho_a}),
replacing also $p$ by $\po(z,t)$.

Notice that another expression for $e_v$
can be derived considering
$\qvstar = \qvstar(q_a,p,T)$ (eq.~(\ref{eqn:qvstar_2}))
in 
$e_v=\rho D \qvstar/Dt$.
Moreover,
two more expressions for
$e_v$ can be 
found using
\begin{equation*}\label{eqn:DTDtehatpres}
\frac{D T}{D t} =
\frac{1}{\rho \cvm} \left[
- p \left(\Dx \cdotb \vel\right) - ( L_e - R_v T )e_v
+ \rho \Hhat
 \right],
\end{equation*}
instead of (\ref{eqn:DTDthhatpres}), 
deduced from 
the conservation equation for 
internal energy $\ehat$ 
instead of enthalpy $\hhat$.
Numerical computations using different formulations for
$e_v$ yield practically identical results.

\section{Modified divergence constraint}\label{app:mod_cons}
Rearranging terms in (\ref{eqn:divu_alpha_dependency}),
after having introduced
the estimate of $e_v$ 
(eq. (\ref{eqn:ev_hhat_rho_a}))
in $S$ (eq. (\ref{eqn:alphapres})), 
yields the modified divergence
constraint (\ref{eqn:divu_ev}) with 
\begin{equation*}
\talpha = \ds
\frac
{\ds \frac{1}{\gamma_m \po} - 
\frac{\Bev}{\rho} 
\left[
\frac{1}{(\epsilon q_a +  q_v)} 
- \frac{L_e}{\cpm T}
\right]}
{1 - \ds
\frac{\Aev}{\rho} 
\left[
\frac{1}{(\epsilon q_a +  q_v)} 
- \frac{L_e}{\cpm T}
\right]
},
\end{equation*}
\begin{equation*}\label{eqn:Stilde}
\tS = \ds
\frac
{\ds \left[ \frac{1}{\cpm T}\right] \Hhat + 
\Cev
\left[
\frac{1}{(\epsilon q_a +  q_v)} 
- \frac{L_e}{\cpm T}
\right]\Hhat }
{1 - \ds
\frac{\Aev}{\rho} 
\left[
\frac{1}{(\epsilon q_a +  q_v)} 
- \frac{L_e}{\cpm T}
\right]
} =
\tsigma \Hhat;
\end{equation*}
and thus
\begin{equation}\label{eq:sigmatilde}
\tsigma =
\frac{\rho
\left\{
(\epsilon q_a +  q_v) + \Cev
\left[
\cpm T - (\epsilon q_a +  q_v)L_e\right]
\right\}
}
{
(\epsilon q_a +  q_v)\rho\cpm T -
\Aev \left[
\cpm T - (\epsilon q_a +  q_v)L_e\right]
}.
\end{equation}
After some manipulation, we can
write that 
$\talpha = 1/(\tgamma \po)$
with
\begin{equation}\label{eqn:gammatilde1}
\tgamma = \gamma_m 
\left[
\ds \frac
{1 + \cvm \Ag}
{1 + R_m T \phi (T) \Ag }
\right],
\qquad
\Ag = 
\ds \frac
{\qvstar \left[
R_m L_e - \cpm R_v T
\right] }
{ \cvm R_m T \left[
\cpm + \qvstar \phi(T) L_e
\right] },
\end{equation}
using (\ref{eqn:ABC1}).
In particular if 
there is no phase transition
($e_v = 0$), 
then $\Ag=0$
and $\tgamma = \gamma_m$;
similarly,
$\Aev=\Cev=0$ in (\ref{eq:sigmatilde}) and 
$\tsigma = \sigma$.

As pointed out in Appendix~\ref{app:evp_rate},
three more expressions for $e_v$, other than 
(\ref{eqn:ev_hhat_rho1}), can be derived
yielding four different formulations for 
$\talpha$ and $\tS$ in the 
modified divergence constraint (\ref{eqn:divu_ev}).
Nevertheless, all of them yield practically identical
numerical results.

\bibliographystyle{alpha}
\bibliography{ws}

\end{document}